\documentclass[final]{siamltex}

\usepackage[]{fontenc}
\usepackage[latin1]{inputenc}
\usepackage{verbatim}
\usepackage{amsmath}
\usepackage{amssymb}
\usepackage{dsfont}
\usepackage{amsfonts}
\usepackage{mathrsfs}
\usepackage{graphicx}
\usepackage{color}
\usepackage{ulem}
\usepackage{pgfplots}
\usepackage{setspace}
\usepackage{xspace}
\usepackage{eclbkbox}
\usepackage{multirow}
\usepackage{ stmaryrd }
\usepackage{graphicx}
\usepackage{lineno}
\usepackage{color}
\usepackage{url}
\usepackage{graphicx}
\usepackage{ stmaryrd }
\usepackage{algorithm}
\usepackage{algpseudocode}
\usepackage{amsmath,amssymb,color} 

\DeclareMathOperator*{\argmin}{arg\,min}
\newcommand\xx{\mathbf{x}}

\newcommand{\Rr}{\mathbb{R}}

\newcommand{\eg}{\textit{e.g.}}
\newcommand{\ie}{\textit{i.e.}}

\newcommand\yy{\mathbf{y}}

\newcommand{\unit}{1\!\!1}
\newcommand{\ctheta}{{\cal{\vartheta}}}

\def\wwr{\mbox{$\mathbf{{v}}$}}
\def\ww{\mbox{$\mathbf{{v}}$}}
\def\tvt{\mbox{$\mathbf{{u}}$}}

\newtheorem{remark}{\it Remark}

\author{P. H\'EAS\thanks{INRIA Rennes \& IRMAR, Universit\'e de Beaulieu, 35042 Rennes, France ({\tt Patrick.Heas@inria.fr}) } \and   F. C\'erou\footnotemark[2]   \and M. ROUSSET\footnotemark[2] 
   }

\begin{document}

\title{ Chilled Sampling for  Uncertainty Quantification: A MOTIVATION FROM A METEOROLOGICAL INverse PROBLEM \thanks{This work was partially supported by the Centre Henri Lebesgue ANR-11-LABX-0020-0.}}

\maketitle
\thispagestyle{empty}

\bigskip

\begin{abstract}
Atmospheric motion vectors (AMVs) extracted from satellite imagery are the only wind observations with good global coverage. They are important features for feeding numerical weather prediction (NWP) models. Several Bayesian models have been proposed to estimate AMVs. Although critical for correct assimilation into NWP models, very few methods provide a thorough characterization of the estimation errors. The difficulty of estimating errors stems from the specificity of the posterior distribution, which is both very high dimensional, and  highly ill-conditioned due to a singular likelihood, which becomes critical in particular in the case of missing data (unobserved pixels). Motivated by this difficult inverse problem, this work studies the evaluation of the (expected) estimation errors  using gradient-based Markov Chain Monte Carlo (MCMC) algorithms. The main contribution is to propose a general strategy, called here  ''chilling'', which amounts to sampling a local approximation of the posterior distribution  in the neighborhood of a point estimate. 
From a theoretical point of view, we show that under regularity assumptions, the family of chilled posterior distributions converges in distribution as temperature decreases to an optimal Gaussian approximation at a point estimate given by the Maximum A Posteriori (MAP), also known as the Laplace approximation.  Chilled sampling therefore provides access to this approximation  generally out of reach in such high-dimensional nonlinear contexts.
From an empirical perspective, we evaluate the proposed approach based on some quantitative Bayesian  criteria. Our numerical simulations are performed on synthetic and real meteorological data. They reveal that not only the proposed chilling exhibits a significant gain in terms of  accuracy of the AMV point estimates  and of their associated expected error estimates, but also a substantial acceleration in the convergence speed of the MCMC algorithms. 
\end{abstract}
\section{Introduction}

The  problem addressed in this paper is a fairly general problem in Bayesian uncertainty quantification. {Assume one tries to estimate a very high dimensional vector-valued parameter, say by computing the maximum a posteriori (MAP) estimate, \ie, maximizing the logarithm of the density (with respect to the flat measure) of an a posteriori distribution. A quantification of the uncertainty of the  estimation may be achieved by using the a posteriori distribution itself, through the characterization of a meaningful norm of the fluctuation around the estimate.} In many cases, state-of-the-art methods will struggle to characterize these fluctuations. For instance, a precise deterministic quadratic approximation around a point estimate of the log-density, often referred to as a Laplace approximation, is generally out of reach in such a context, due to the poor numerical approximation of the Hessian matrix of the posterior log-density, and the prohibitive computational complexity of its inversion. On the other hand, state-of-the-art sampling of the a posteriori distribution often yields poor results, due to extremely slow convergence in high dimensions, especially when the posterior log-density is poorly conditioned and exhibits {multi-scale features with non-Gaussian or even multi-modal issues.} \medskip

{The general idea of the present work will be to use well-tuned (preconditioned) Markov chain Monte Carlo (MCMC) methods for sampling \textit{Gaussian local approximation} around a point estimate of the posterior distribution; as opposed to the full posterior distribution.}\medskip 

This general uncertainty quantification method has been designed while studying a specific practical application in meteorology {which exhibits the difficulties outlined above}. We will thus propose our general Bayesian methodology through this specific application, and use the latter as a large-scale real-world example to demonstrate the potential of our approach. \medskip

The considered specific problem constitutes a critical issue in data assimilation and weather forecasting. Let us describe briefly this specific problem and its context.
The  improvement  of  numerical  weather  prediction  (NWP)  forecast  models  requires  the assimilation of meteorological observations that inform the models about the current state of  the  atmosphere.  Therefore,  NWP  models  are  continuously  fed  with  a  wide  range  of  in-situ  observations,  like  radiosondes,  radars,  buoys,  aircraft  measurements,  and  observations  extracted  from satellite data. The proportion of satellite data assimilated in NWP models has increased a lot during recent years because these models cover all areas around the Earth, including the oceans and the  polar  regions,  where  few  in-situ  measurements  are  available~\cite{borde2019winds}.  Atmospheric  motion  vectors  (AMVs) derived by tracking clouds or water vapour features in consecutive satellite images, constitute the only wind observations with good global coverage that help to predict the evolution and displacement of air masses.  
Recent  studies  have  been  conducted  to  investigate  the  extraction  of  AMV  profiles  from  moisture  and  temperature  fields  retrieved  from  hyperspectral  instruments~\cite{borde2019winds,santek2019demonstration,Heas_2023}. These studies have pointed out that energy minimization methods, known in the computer vision literature as {\it optic-flow} algorithms, are promising approaches in atmospheric sciences because of their good adaptation to the inherent physical  nature of the images, and because they can deal with low contrasted and missing observations.   
Thorough error characterization is critical in properly assimilating AMVs into NWP models. Since it is known that portions of the AMVs are unreliable, the density of the vector fields is reduced before assimilation.
A common practice in the atmospheric sciences community  is to filter out the AMVs which are assigned to a so-called low {\it quality indicator}~\cite{santek20192018}. 
These indicators are based on comparing changes in AMV estimates between sequential time steps and neighboring pixels, as well as differences with model predictions.  Other approaches build statistical model using linear regression against 
radiosonde values to correct AMV observation error~\cite{le2004error} or rely on machine learning techniques and training data generated by independent NWP simulations~\cite{teixeira2021using}. 
On the other hand, very few   optic-flow  methods   propose  uncertainty estimates  in the computer vision literature.
Post-hoc methods apply post-processing to already estimated flow fields~\cite{kondermann2007adaptive,kondermann2008statistical,mac2012learning}.
Other methods, in contrast, produce their uncertainty estimates relying on approximations of {a Bayesian posterior distribution associated to} the  optic-flow estimation problem. 
Bootstrap sampling   on the data term is proposed in \cite{kybic2011bootstrap}. Gibbs sampling of a linearized posterior model  is devised in~\cite{sun2018bayesian}. The authors in \cite{wannenwetsch2017probflow} derive a mean-field approximation of the  posterior distribution 
 and  provide a computationally efficient uncertainty estimation method using variational Bayes. Besides, learning strategies based on deep networks have recently emerged  for optic-flow uncertainty estimation~\cite{ummenhofer2017demon, ilg2018uncertainty}. \medskip

Returning to the general level of the proposed methodology, which aims to mitigate the poor behavior of state-of-the-art sampling techniques, the idea is to use gradient-based Markov chain Monte Carlo (MCMC) sampling methods, but to sample a (local) Gaussian approximation  of the posterior distribution around a point estimate.  
To do so, we propose a ``chilling'' strategy which enforces the sampling of a local approximation of the posterior in the neighborhood of a given point estimate of the MAP. More explicitly,  for a posterior distribution of the form $\mu(d\boldsymbol{\theta}) \propto \exp^{-U(\boldsymbol{ \theta})} {d\boldsymbol{\theta}}$, with $U$ precisely defining the posterior log-density and   ${d\boldsymbol{\theta}}$ a reference Lebesgue (flat) distribution used in the definition of the prior, the chilled posterior distribution at temperature $\zeta \in (0,1]$ in the neighborhood of a point estimate  $\boldsymbol{\hat \theta}$ of the MAP will be defined  by 
$$
\mu_{\zeta}(d\boldsymbol{\theta}) \propto {\exp^{-\frac{1}{\zeta}U(\boldsymbol{\hat \theta}+\zeta^{1/2}(\boldsymbol{\theta}-\boldsymbol{\hat \theta})) }\, {d\boldsymbol{\theta}}},
$$
 as exposed in Section~\ref{sec:contributions}. 
 We will focus on Gaussian priors, although the proposed methodology is not limited to this particular case.
Under regularity assumptions, and taking $\boldsymbol{\hat \theta}=\int \boldsymbol{ \theta} d  \mu_{\zeta}(d\boldsymbol{\theta})$, our first contribution is to show that the family of chilled posterior distributions converges, as temperature decreases, towards the {optimal} Gaussian approximation around the MAP point estimate, this limiting approximation being the Laplace approximation\footnote{We point out that the proposed chilling can be seen to some extent  as the antipode of {simulated annealing}~\cite{van1987simulated}:   the simulation in the latter starts at a higher temperature while in the former it starts and remains at lower temperature.  }. Our approach then relies on Markov Chain Monte-Carlo sampling, but because the target distribution is an approximate Gaussian distribution concentrated in a very small volume, {it is amenable to preconditioning (\ie, an initial guess of the practice-dependent covariance can be injected into the Markov chain setting}), and is then expected to converge much faster. Being focused on algorithmic efficiency, the paper concentrates on gradient-based MCMC simulations of the chilled distributions, in particular the  Metropolis-Adjusted Langevin Algorithm (MALA) and  Hamiltonian Monte Carlo (HMC), which are known to be state-of-the-art sampling methods.\medskip

A second contribution of this work  concerns the application of the chilled sampling method to the specific meteorological example of interest. The chilled sampling is tuned using an {efficient preconditioning} of gradient-based MCMC simulations. The preconditioning is based on the prior family itself but with possibly differently adjusted hyper-parameters. 
More precisely, the prior distribution considered in this work is given by a two-dimensional isotropic fractional Brownian motion (fBm) with two hyper-parameters (point variance and Hurst coefficient) that are tuned in advance using expert knowledge. 
\medskip


We focus in our numerical simulations on evaluating, on the one hand, the precision of the AMV estimates with respect to the ground truth and, on the other hand, their uncertainty estimates.   In the perspective of integrating precision and uncertainty in a single measure, we propose several evaluation criteria, which are optimal in a certain sense. They take the form of a weighted average of the errors, where the optimal weights are functions of the error estimates.     Using these criteria, we are specifically interested in the short-term trend toward convergence of MCMC algorithms, as the simulation budget available in a meteorological operational procedure is insufficient to achieve full convergence. Since the MAP estimate computed by deterministic optic flow procedures generally differs from the optimal Bayesian estimate, a budget-constrained MCMC simulation starting from such a deterministic estimate often remains dependent on the initial condition.\medskip

In addition to an increase in  estimation accuracy of the point-wise  mean and variance, {our criteria indicate} that there is a significant gain (as temperature decreases) in terms of the convergence speed of MCMC algorithms. Since the {proposed chilling leaves the quadratic approximation of the log-density posterior invariant}, this acceleration is caused exclusively by the attenuation of nonlinearities in the gradient of the log-density posterior (presumably in the slower directions).
On the basis of our experiments, we conjecture that the relevance of the proposed chilled approximation of the posterior distribution goes beyond the case of (almost) Gaussian posteriors, although the determination of sufficient conditions guaranteeing the quality of this approximation remains an open problem.\medskip

%
%

The paper is organized as follows. In section~\ref{sec:4}, we first define the Bayesian context and objectives of uncertainty quantification, but also the criteria for evaluating the quality of error estimates. We then  
 introduce our chilled approximations of the posterior in Section~\ref{sec:chillingTheory} and study their properties. Simulation of the chilled posterior  by gradient-based MCMC algorithms to achieve effective uncertainty quantization is discussed in Section~\ref{sec:MCMCgrad}. In Section~\ref{Sec:Model}, we present the specific Bayesian modeling adopted for the specific inverse problem of interest: AMV estimation in the context of partial image observations.  In Section~\ref{sec:numEval}, we numerically evaluate the proposed algorithms on synthetic and real weather data.  Finally, the appendix collects some technical proofs and details about the algorithms.

    \section{General Framework and Objectives}\label{sec:4}
  \subsection{A Bayesian model}

We adopt a Bayesian formulation of the problem of interest. It assumes that both, the unknown ground truth $\boldsymbol{\theta}^\star \in \mathbb{R}^n$ and the observations $\yy \in \mathcal{Y}$, are  realizations of  random variables.
  For reasons of presentation, we report the specific posterior model used for AMVs and partial image observations in section~\ref{Sec:Model}, and limit ourselves to presenting a generic modeling framework below. 
  
We assume a differentiable function $\varphi: \Rr^n\to \Rr $ defining a prior distribution $\nu_0$ for the random variable $\boldsymbol{\theta}$ of the form  
 $$\varphi(\boldsymbol{\theta})=-\log \nu_0(\boldsymbol{\theta}).$$
 Gaussian priors will be of  interest in our numerical simulations, although the proposed methodology does not restrict to this particular case.  
 For a centered multivariate Gaussian  $\mathcal{N}(0,\boldsymbol{\Sigma}_0)$ with $\boldsymbol{\Sigma}_0 \in \Rr^{n \times n}$,  $$\varphi(\boldsymbol{\theta})= \boldsymbol{\theta}^\intercal\boldsymbol{\Sigma}_0^{-1}\boldsymbol{\theta}.$$
The covariance matrix $\boldsymbol{\Sigma}_0$  will have  a  structure specific to the statistics of the  field of interest. 
Denoting the negative log-likelihood by $\phi: \Rr^n\times \mathcal{Y} \to \Rr$, by Bayes' theorem, we get  the posterior
\begin{align}\label{eq:posterior}
\mu^\yy(d{\boldsymbol{\theta}})
=\frac{1}{Z}\exp^{-U(\boldsymbol{\theta};\yy)}d{\boldsymbol{\theta}},
\end{align}
where we have defined the Gibbs energy $U(\boldsymbol{\theta};\yy)=\phi(\boldsymbol{\theta};\yy)+\varphi(\boldsymbol{\theta})$  and its normalization constant $Z=\int \exp^{-U(\boldsymbol{\theta};\yy)}d{\boldsymbol{\theta}}$.
A Bayesian estimator is then defined as 
\begin{align}\label{eq:BayesEstim}
  \boldsymbol{\theta}_\mathcal{L}\in \underset{ \boldsymbol{\theta}'\in \Theta}{\arg \min}\,\int  \mathcal{L}(\boldsymbol{\theta},\boldsymbol{\theta}'){ \mu^{\yy}({\boldsymbol{\theta}})}d{\boldsymbol{\theta}},
\end{align}
where $\mathcal{L}:\Theta \times \Theta\to \Rr$ is a  cost function.  In particular,  the cost $\mathcal{L}(\boldsymbol{\theta},\boldsymbol{\theta}')=1-\delta(\boldsymbol{\theta},\boldsymbol{\theta}')$  yields  the maximum a posteriori~(MAP) 
\begin{align}\label{objFunction5pre}
 \boldsymbol{\theta}_{MAP} \in \underset{ \boldsymbol{\theta}\in \Theta}{\arg \min} \,U(\boldsymbol{\theta};\yy).
\end{align}
Alternatively,  a quadratic cost $\mathcal{L}(\boldsymbol{\theta},\boldsymbol{\theta}')=(\boldsymbol{\theta}-\boldsymbol{\theta}')^\intercal(\boldsymbol{\theta}-\boldsymbol{\theta}')$ leads  to the posterior mean (PM) \vspace{-0.3cm}
\begin{align}\label{eq:PMestim}
  \boldsymbol{\theta}_{PM}=\int \boldsymbol{\theta} { \mu^{\yy}({\boldsymbol{\theta}})}d{\boldsymbol{\theta}} .
\end{align}
MCMC algorithms are popular to compute  and estimate  the PM  \eqref{eq:PMestim}. However, they are in general much more computational demanding  than the the standard efficient techniques used for MAP estimation. 

    For simplicity we drop the notation $\yy$ in the following from the various terms involved, so we denote the posterior as  $\mu$.

\subsection{The objective: evaluating the expected error norm}\label{sec:eQerror}
  Function $\psi:\Rr^n\to \Rr^\ell$, $\ell \in \mathbb{N}_+$ will  denote in the following bounded linear operators. These linear test functions $\psi$ belong to some  given finite set $\mathcal{P}(\Omega)$. We are interested in quantifying  the uncertainty of the estimate $\psi( \boldsymbol{\hat \theta})$, where $ \boldsymbol{\hat \theta}$ is an estimate of $ \boldsymbol{\theta}_\mathcal{L}$, which accounts to characterize the distribution of the  error between the estimate $ \psi(\boldsymbol{\hat \theta})$ and the unknown ground truth $\psi( \boldsymbol{ \theta}^\star)$. In the following, we will restrict ourselves to the evaluation of the Euclidian error norm. 
  
As we have assumed that $\boldsymbol{\theta}^\star$ was drawn according to $\mu$,  the expectation of the error norm  $\|\psi(\boldsymbol{\theta}^\star)-\psi( \boldsymbol{\hat \theta} )\|_2$ is  given  by 
   \begin{align}\label{eq:varianceExpectedError}
  {\mathcal{E}}_{\mu,\boldsymbol{\hat \theta}}(\psi) =\int \|\psi(\boldsymbol{\theta})-\psi( \boldsymbol{\hat \theta} )\|_2\, \mu(d{\boldsymbol{\theta}}),
  \end{align}
Alternatively,  the Chebyshev inequality  characterizes the probability that the  error $\|\psi(\boldsymbol{\theta}^\star)-\psi( \boldsymbol{\hat \theta} )\|_2$ exceeds a given threshold value
  \begin{align}\label{eq:boundProba}
 {\mathbb{P}_{\mu} \left( \|\psi(\boldsymbol{\theta})-\psi( \boldsymbol{\hat \theta} )\|_2 \ge a \right)}&\le \frac{ {\mathcal{E}}_{\mu,\boldsymbol{\hat \theta}}(\psi)}{a},\quad a\in \Rr_+.
  \end{align}
Uncertainty quantification consists in both cases, \eqref{eq:varianceExpectedError} and \eqref{eq:boundProba}, in evaluating the expectation ${\mathcal{E}}_{\mu,\boldsymbol{\hat \theta}}(\psi) $.    \\
  
\subsection{An assessment criteria: the averaged weighted error norm}

In an evaluation context,  we generally have at our disposal  the true parameter $\boldsymbol{ \theta}^\star$, which has generated  the observation $\mathbf{y}$, \eg, in the context of  synthetic data provided by a NWP model simulation. \\
We propose a criteria to jointly assess the accuracy of the estimate $\boldsymbol{\hat \theta}$ and the performance of the uncertainty quantification procedures proposed in the next sections. 
For the meteorological problem under consideration, the $\boldsymbol{\hat \theta}$ components will be the bi-variate AMV estimates associated with the spatial coordinates of the two-dimensional domain $\Omega$, while the errors will be the AMV endpoint errors  (EPE).  The norm of these errors will be weighted by a function ${w}:\mathcal{P}(\Omega) \to \Rr_+$, designed as a non-increasing function of the expected error ${\mathcal{E}}_{\mu,\boldsymbol{\hat \theta}}(\psi)$.  Averaging these weighted error norms over the linear test functions $\psi \in \mathcal{P}(\Omega)$  will lead to the following assessment criterion
\begin{equation}\label{eq:MSE}
\textrm{EPE}(\Omega,{w},\boldsymbol{\theta^\star}-  \boldsymbol{\hat \theta})=\frac{1}{\sharp\mathcal{P}(\Omega)}\sum_{{\psi} \in \mathcal{P}(\Omega)} {w}(\psi) \|\psi(\boldsymbol{\theta^\star}-   \boldsymbol{\hat \theta} )\|_2.
\end{equation} 
 For instance,  one may define a weighting function inversely proportional to a power $p>0$ of the expected error
\begin{equation}\label{eq:weightL1_}
{w}_{\mu,\boldsymbol{\hat \theta}}^{p}(\psi) ={c}_p {  {{\mathcal{E}}_{\mu,\boldsymbol{\hat \theta}}(\psi)}^{-p}},
\end{equation}
or consider binary weights \vspace{-0.2cm}
 \begin{align}\label{eq:weightL0_}
{w}_{\mu,\boldsymbol{\hat \theta}}^{0}(\psi)=
 \left\{\begin{aligned}
&{c}_0 \quad \textrm{if} \quad{  {{\mathcal{E}}_{\mu,\boldsymbol{\hat \theta}}(\psi)}}  \le  \textrm{d}_0\\
&0\quad \textrm{else} 
\end{aligned}\right.,
\end{align}
with  positive constants  ${c}_0$, ${c}_p$ and ${d}_0$. Note that function \eqref{eq:weightL0_} accounts to restrict to the sub-set of  estimates for which the probability that the  error exceeds a certain value is lower than a bound of the form of~\eqref{eq:boundProba}. Besides this possible interpretation, we can show that functions \eqref{eq:weightL1_} and \eqref{eq:weightL0_} are optimal in some sense. However, before detailing this property, we  introduce a lemma  guaranteeing that, for any approximation ${\hat \mu}$ of the posterior $\mu$ and any estimate $\boldsymbol{\hat \theta}$,  the expectation of the weighted average error $\eqref{eq:MSE}$ is an upper bound on the expected weighted average error around the posterior mean, with weights computed with the true posterior. The inequality tends to become an equality when $\boldsymbol{\hat \theta} \to  \boldsymbol{\theta}_{PM}$ and $\hat \mu \to \mu$. \\
 
\begin{lemma}\label{lemma:approxsol}
For any estimate $\boldsymbol{\hat \theta}$ and posterior approximation $\hat \mu$, the following inequality holds 
$$\int\textrm{EPE}(\Omega,{w}_{\hat \mu,\boldsymbol{\hat \theta}}^{p},\boldsymbol{\theta^\star}-  \boldsymbol{\hat \theta}) \mu(d\boldsymbol{\theta}^\star) \ge \int \textrm{EPE}(\Omega,{w}_{\mu,\boldsymbol{\theta}_{PM}}^{p},\boldsymbol{\theta^\star}- \boldsymbol{\theta}_{PM}) \mu(d\boldsymbol{\theta}^\star),$$
where the weighting functions are  \eqref{eq:weightL1_} or \eqref{eq:weightL0_} substituting  $\hat \mu$ for $\mu$ or $\boldsymbol{\hat \theta}$  for  $\boldsymbol{\theta}_{PM}$. \\
\end{lemma}
 
  The proof (given in Appendix~\ref{app:proofLemma}) relies on the fact that $\boldsymbol{\theta}^\star$ is  a random variable distributed according to the posterior law and on the optimality of function ${w}_{\mu,\boldsymbol{\theta}_{PM}}^{p}$, which we now discuss.
Consider the  optimization problem
\begin{equation}\label{eq:minProb}
{w}_{\mu,\boldsymbol{\hat \theta}}^{p}= \argmin_{{w}\in \mathcal{W}}\int \textrm{EPE}(\Omega,{w},\boldsymbol{\theta^\star}-  \boldsymbol{\hat \theta}) \mu(d\boldsymbol{\theta}^\star)\quad \textrm{s.t.}\quad h_p({w})=0,
\end{equation}
 given $h_p: \mathcal{P} \to \Rr$, which  specifies some scalar constraint, and given some admissible set  $\mathcal{W}$.
As shown  in  the appendix, the weighting functions \eqref{eq:weightL1_} and \eqref{eq:weightL0_}  correspond indeed  to unique minimizers of  \eqref{eq:minProb}.  In particular, function \eqref{eq:weightL1_} where  $p=1$ and ${c}_1={\prod_{{\psi}\in \mathcal{P}}{  {{\mathcal{E}}_{\mu,\boldsymbol{\hat \theta}}(\psi)}^{{1/{\sharp\mathcal{P}(\Omega)}}}} }$  correspond to the optimal solution for the constraints
\begin{align}\label{eq:const0}
h_1({w})=\sum_{{\psi} \in \mathcal{P}(\Omega)} -\log({w}(\psi)) \quad \textrm{and}\quad \mathcal{W}=\{{w}: \mathcal{P} \to \Rr_+\}.
\end{align}
Function  \eqref{eq:weightL1_}  where  $p=2$ and ${c}_2={{\sharp\mathcal{P}(\Omega)}}^2({\sum_{{\psi}\in \mathcal{P}}{  {{\mathcal{E}}_{\mu,\boldsymbol{\hat \theta}}(\psi)}^{-1}}})^{-2}$ is the optimal solution  related to the constraints  
\begin{align}\label{eq:const1}
h_2({w})= { {\sharp\mathcal{P}(\Omega)}}-\sum_{{\psi} \in \mathcal{P}(\Omega)} \sqrt{{w}(\psi) } \quad \textrm{and}\quad \mathcal{W}=\{{w}: \mathcal{P} \to \Rr_+\}.
\end{align}
Finally function \eqref{eq:weightL0_}, where ${c}_0={{\sharp\mathcal{P}(\Omega)}/\tau}$ and ${d}_0\in [0,{\sum_{{\psi'}\in \mathcal{P}}{  {{\mathcal{E}}_{\mu,\boldsymbol{\hat \theta}}(\psi)}}}] $ depends on $\tau\in (0,{\sharp\mathcal{P}(\Omega)}]$, is the optimal solution satisfying the constraints
\begin{equation}\label{eq:const2}
h_0({w})=\tau- \sum_{{\psi} \in \mathcal{P}(\Omega)} \unit_{\{{w}(\psi)>0 \}}  \quad\textrm{and}\quad \mathcal{W}=\{{w}: \mathcal{P} \to \{0,{{\sharp\mathcal{P}(\Omega)}/\tau}\}\}.
\end{equation}

  \section{Chilled distributions and  Laplace approximation}\label{sec:chillingTheory}
The practical computation of the expectation ${\mathcal{E}}_{\mu,\boldsymbol{\hat \theta}}(\psi) $ is very demanding in our high dimensional environment. On the one hand, this quantity is analytically or numerically difficult to calculate  
 because the posterior distribution
is not available in a closed form. On the other hand, the application of straightforward Monte-Carlo (MC) approximations fails in such high dimensional scenarios, as illustrated by our numerical experiments in Section~\ref{sec:numEval}.
The evaluation of the expected error will  typically rely on substituting  the posterior $\mu$ by some  approximation. 
   
\subsection{Laplace method: a point-wise Gaussian approximation}\label{sec:laplace}
A common strategy  is to resort to a Gaussian assumption for $\mu$ and  building a so-called Laplace  approximation of the expectation ${\mathcal{E}}_{\mu,\boldsymbol{\hat \theta}}(\psi) $.  As we will see hereafter, the Laplace approximation provides an exact   evaluation of this quantity as long as the Gaussian assumption holds.  As shown in the appendix, its calculation is tractable for some posterior structures. It should be noted, however, that tractability often requires coarse assumptions and numerical approximations, resulting in poor Laplace approximations, as illustrated by our numerical experiments in Section~\ref{sec:numEval}. Nevertheless, we will present the Laplace approximation below, as it will provide a reference for later use in Section~\ref{sec:contributions}.   
 
 More precisely,  assume   the log-posterior  is twice differentiable.    As proposed in~\cite{berger2013statistical}, it is convenient to approximate  the  posterior  $\mu$
 by the normal law 
 \begin{align}\label{eq:mu0}
 \mu_0=\mathcal{N}( \boldsymbol{\hat \theta},{\mathbf{H}_{U}^{-1}}),
 \end{align}  with the   covariance matrix  ${\mathbf{H}_{U}^{-1}}$  given as  the inverse  of the Hessian (which we will assume positive definite) of the log posterior
 \begin{align}\label{eq:invExpectLaplace}
 \mathbf{H}_{U}(i,j)
 = \frac{\partial^2  U(\boldsymbol{\theta})}{\partial \boldsymbol{\theta}(i)\partial \boldsymbol{\theta}(j)}\Bigl\vert_{ \boldsymbol{\theta}=\boldsymbol{\hat \theta}}.
  \end{align}
  This  is known as the Laplace approximation.
  We remark that under the normal assumption, the components of  $\mathbf{H}_{U}$  are assumed to be invariant with respect to $\boldsymbol{\theta}$. As we will illustrate in our numerical simulations, this invariance assumption will constitute a  limitation of the Laplace methodology. 

 Under the Gaussian assumption, the expected  error ${\mathcal{E}}_{ \mu,\boldsymbol{\theta}_{MAP}}(\psi) $ is explicit.  To avoid the dependance of  ${\mathcal{E}}_{ \mu,\boldsymbol{\theta}_{MAP}}(\psi) $ to the correlation of  the components of $\psi(\boldsymbol{\theta})$, we propose to bound the expected error using   norm equivalences.    The bound  characterization will necessitate  the  eigenvalue decomposition (EVD)  of  the symmetric   matrix ${\mathbf{H}_{U}}$. We write in matrix form the EVD as  ${\mathbf{H}_{U}}=\mathbf{V}_{{\mathbf{H}_{U}}} \Lambda_{\mathbf{H}_{U}} \mathbf{V}_{{\mathbf{H}_{U}}}^\intercal $. \\

\begin{remark}\label{Prop:1}
Since function  $\psi:\Rr^n\to \Rr^\ell$ satisfies $$ \frac{1}{\sqrt{\ell}}  \|\psi(\boldsymbol{\theta} )\|_1\le \|\psi(\boldsymbol{\theta}) \|_2 \le \|\psi(\boldsymbol{\theta}) \|_1, $$  under a Gaussian assumption it is straightforward to show that the expected  error \eqref{eq:varianceExpectedError} is bounded as
\begin{align}\label{eq:boundEllipsRestrict}
\frac{ \mathcal{F}}{\sqrt{\ell}} \le {\mathcal{E}}_{ \mu,\boldsymbol{\theta}_{MAP}}(\psi)   \le  \mathcal{F} \quad \textrm{with} \quad \mathcal{F}=\sqrt{\frac{2}{\pi}}   \sum_{j=1}^\ell \| \Lambda_{{\mathbf{H}_{U}}}^{-1/2} \mathbf{V}_{{\mathbf{H}_{U}}}^\intercal   {\mathbf{\psi}_j}\|_2,
\end{align}
with the $j$-th component of $\psi(\boldsymbol{\theta})$    denoted by  $\mathbf{\psi}_j^\intercal\boldsymbol{\theta}$ for some $\mathbf{\psi}_j\in \Rr^n$.\\

\end{remark}
%

 The computation of the  EVD of the inverse covariance matrix required in \eqref{eq:boundEllipsRestrict}  is at first glance intractable because it requires a  complexity of~$\mathcal{O}(n^3)$, which is generally  prohibitive  for very large   $n$. However, we show in the appendix that the complexity may be significantly reduced, in the case of  of particular posterior distributions, by exploiting conditional independence.  

However, the Laplace approximation often performs poorly, as it is shown  in our numerical experiments presented in Section~\ref{sec:numEval}.  Indeed, the Laplace method is a point estimate at $\boldsymbol{\hat \theta}$ of the Hessian of the log posterior density.   The latter is strongly ill-conditioned due to the multi-scale nature of the high-dimensional problem. In addition, the shortest scales of the logarithmic density can be irregular in the neighborhood of $\boldsymbol{\hat \theta}$, exhibiting multiple local maxima and strong nonlinearities. These factors probably contribute to the lack of robustness of the Laplace approximation that we observed numerically.

 \subsection{Chilling: a family of local posterior approximations}\label{sec:contributions}
To avoid a  restrictive Laplace approximation while keeping  computation time reasonable,  another track  is to rely on chilling. By introduction of a  low-temperature parameter $\zeta \in (0,1]$, we define a continuous  family of probabilistic laws, ranging  from  the original posterior to  the Gaussian law used in the Laplace method. 

More precisely, 
 we introduce for   $\zeta \in (0,1]$ the (non-rescaled) {\it chilled}  distribution \vspace{-0.cm}
\begin{equation} \label{eq.pizeta}
\boldsymbol{\tilde \ctheta} \sim \pi_\zeta(d\boldsymbol{\tilde \theta})\propto {\exp(- U_\zeta(\boldsymbol{\tilde \theta)) }\, d\boldsymbol{\tilde \theta}},
\end{equation}
with the chilled  energy $U_\zeta(\boldsymbol{\tilde \theta})=\zeta^{-1}U(\boldsymbol{\tilde \theta})$.  We remark that the mass of the density of random variable  $\boldsymbol{\tilde \ctheta}$ tends to concentrate  on the MAP as $\zeta$ decreases. We associate with $\boldsymbol{\tilde \ctheta}$, a rescaled random variable
$\zeta^{-1/2}(\boldsymbol{\tilde \ctheta}-\boldsymbol{\hat \theta})$
  centered on $\boldsymbol{\hat \theta}$, whose distribution will be asymptotically non trivial as $\zeta\rightarrow 0$.   We will set 
  \begin{equation}\label{eq:ChilledPostMean}
  \boldsymbol{\hat \theta}=\int \boldsymbol{\tilde \theta} d  \pi_\zeta(d\boldsymbol{\tilde \theta}).
  \end{equation}
Finally,  by a translation of $\boldsymbol{\hat \theta}$, we propose to approximate the posterior  by sampling the random variable 
 \begin{align}
\boldsymbol{ \ctheta}=\boldsymbol{\hat \theta}+\zeta^{-1/2}(\boldsymbol{\tilde \ctheta}-\boldsymbol{\hat \theta}),\label{eq:changevar2}
\end{align} 
 following the (rescaled) chilled posterior distribution
  \begin{align}\label{eq:paramDistrib}
\mu_\zeta(d\boldsymbol{\theta})=
&Z_\zeta^{-1}{\exp(-U_\zeta(\boldsymbol{\hat \theta}+\zeta^{1/2}(\boldsymbol{\theta}-\boldsymbol{\hat \theta})) )\, {d\boldsymbol{\theta}}},\quad \textrm{for} \quad \zeta \in (0,1],
\end{align}
 of  normalization constant  $Z_\zeta=\int {\exp(-U_\zeta(\boldsymbol{\hat \theta}+\zeta^{1/2}(\boldsymbol{\theta}-\boldsymbol{\hat \theta})) ){d\boldsymbol{\theta}}}$. 
At unit temperature it thus corresponds to the original posterior, and on the other hand, at a null temperature, we define $\mu_0(d\boldsymbol{\theta})$ as the Laplace approximation \eqref{eq:mu0}.
 To sum up, our approach here is to sample from $\pi_\zeta$ given in equation (\ref{eq.pizeta}), and then use (\ref{eq:changevar2}) to get a sample approximating the posterior. 
 
 The following proposition shows that the family  \eqref{eq:paramDistrib}  (with \eqref{eq:mu0} when $\zeta=0$) is continuous on $\zeta \in [0,1]$. Indeed,  lowering temperature amounts to get closer to a local Gaussian approximation of the posterior around  $\boldsymbol{\hat \theta}$, and  in the limit of a zero temperature, the variable of law~\eqref{eq:paramDistrib}  converges in distribution to the  point-wise Laplace approximation \eqref{eq:mu0}.
 

 For every multi-index $\mathbf{p}=(p_1,\dots,p_n)\in\mathbb{N}^n$ we denote $|\boldsymbol{\theta}|^\mathbf{p}= \prod_{j=1}^d |\boldsymbol{\theta}(j)|^{p_j}$. \\

\begin{proposition}\label{remark:gaussianApprox}
Let $\mu_\zeta$ be defined by \eqref{eq:ChilledPostMean} and  \eqref{eq:paramDistrib}. Assume there exists a postive definite matrix  ${\mathbf{Q}} \in \Rr^{n \times n}$ s.t.
\begin{align*}
U(\boldsymbol{ \theta})-U({\boldsymbol{ \theta}_{MAP}})\ge (\boldsymbol{ \theta}-{\boldsymbol{ \theta}_{MAP}})^\intercal{\mathbf{Q}} (\boldsymbol{ \theta}-{\boldsymbol{ \theta}_{MAP}}), \quad \forall \boldsymbol{ \theta}\in \Rr^n.
\end{align*}
 Then,  if the log posterior   is of class $\mathcal{C}^3(\mathbb{R}^n)$ with uniformly bounded third-order derivatives, when temperature $\zeta$ tends to zero, the convergence in distribution to the Laplace approximation holds  \vspace{-0.2cm}
 $$\boldsymbol{ \ctheta} \sim \mu_\zeta \overset{d}{\to} \mathcal{N}( {{\boldsymbol{ \theta}_{MAP}}},{\mathbf{H}_{U}^{-1}}),$$
where $\boldsymbol{ \ctheta}$  and $\mu_\zeta$ are given by equations (\ref{eq:changevar2}) and  (\ref{eq:paramDistrib}), and in terms of   moments we have
 \begin{align*}
\lim_{\zeta \to 0} \int|\boldsymbol{ \theta}|^{\mathbf{p}} \mu_\zeta(d\boldsymbol{ \theta})=  \int |\boldsymbol{ \theta}|^\mathbf{p}\frac{\exp(-\frac{1}{2}(\boldsymbol{ \theta}-{\boldsymbol{ \theta}_{MAP}} )^\intercal{\mathbf{H}_{U}}(\boldsymbol{ \theta}-{\boldsymbol{\theta}_{MAP}} )) }{(2\pi)^{n/2}|\det {\mathbf{H}_{U}}|^{-1/2}}d\boldsymbol{ \theta},\quad \forall \mathbf{p}\in\mathbb{N}^n.
 \end{align*}
\end{proposition} 

\proof{
We will first consider the case, slightly different, where $\boldsymbol{\hat\theta}$ is replaced by $\boldsymbol{\theta}_{MAP}$ in the definition of the rescaled variable \eqref{eq:changevar2}. 
Since $\boldsymbol{ \theta}_{MAP} $ is by definition a stationary point, the first order derivative of the log posterior vanishes, and a second order Taylor-Lagrange expansion of $U\in \mathcal{C}^3(\Rr^n)$ with  the  multi-index integer notation $\boldsymbol{\alpha}$  gives $\forall \boldsymbol{\tilde\theta}\in \Rr^n$,
\begin{align*}
U(\boldsymbol{\tilde \theta})=U(\boldsymbol{ \theta}_{MAP})+\frac{1}{2} (\boldsymbol{\tilde \theta}-\boldsymbol{ \theta}_{MAP})^\intercal {\mathbf{H}_{U}} (\boldsymbol{\tilde \theta}-\boldsymbol{ \theta}_{MAP})+ \sum_{|\alpha|=3} R_{\boldsymbol{\alpha}}(\boldsymbol{\tilde\theta}) (\boldsymbol{\tilde\theta}-\boldsymbol{ \theta}_{MAP})^{\boldsymbol{\alpha}},
\end{align*} 
 where the matrix  ${\mathbf{H}_{U}}$ is defined in~\eqref{eq:invExpectLaplace} and the remainder satisfies
\begin{align}\label{eq:RemBound}
|R_{\boldsymbol{\alpha}}(\boldsymbol{\tilde \theta})|\le \frac{1}{6}\sup_{|{\boldsymbol{\beta}}|=|{\boldsymbol{\alpha}}|}\sup_{\boldsymbol{\tilde\theta}'\in \Rr^n}|\frac{ \partial^{\boldsymbol{\beta}} U(\boldsymbol{\theta}')}{\partial \boldsymbol{\theta}^{\boldsymbol{\beta}}}| < \infty,\quad \forall \boldsymbol{\tilde\theta}\in \Rr^n,
\end{align}
the last inequality being due to the uniformly bounded third-order derivatives of $U$ on $\Rr^n$.
Rewriting  the Taylor expansion  with respect to the variable 
\begin{align}\label{eq:changeVar}
\mathbf{h}=\zeta^{-1/2}(\boldsymbol{\tilde \theta}-\boldsymbol{\theta}_{MAP}),
\end{align}
 we obtain 
\begin{align*}
U_\zeta(\mathbf{ h})=U_\zeta(0)+ \frac{1}{2 }\mathbf{ h}^\intercal{\mathbf{H}_{U}}  \mathbf{ h}+ \zeta^{1/2}\sum_{|\alpha|=3} R_\alpha(\boldsymbol{ \theta}_{MAP}+\mathbf{h}\zeta^{1/2})\mathbf{ h}^\alpha,\quad \forall \mathbf{ h}\in \Rr^n. \vspace{-0.cm}
\end{align*} 
We will show a result slightly stronger than the convergence in distribution, by considering for the set of test functions those functions $\psi$ that are continuous, and such that $|\psi|$ can be bounded by a polynomial function, i.e. there is a polynomial function $\mathcal{P}$ such that for all $\mathbf{ h}$, $|\psi( \mathbf{ h})| \leq \mathcal{P}(\mathbf{ h})$.

By taking the zero-temperature  limit, it follows from \eqref{eq:RemBound} that for any  $\psi$ 
we have the point-wise convergence\vspace{-0.2cm}
\begin{align}\label{eq:point-wiseCV}
\lim_{\zeta \to 0}  \exp\{ U_\zeta(0)\}\psi(\mathbf{h}) \exp\{-(U_\zeta(\mathbf{ h}))\}
=\psi(\mathbf{h}){\exp\{ -\frac{1}{2}\mathbf{h}^\intercal{\mathbf{H}_{U}} \mathbf{h}} \}.
\end{align}
In addition,  the assumption leads to the existence of a positive definite matrix  ${\mathbf{Q}} \in \Rr^{n \times n}$ such that $\forall \mathbf{ h}\in \Rr^n$\vspace{-0.2cm}
\begin{align*}
U_\zeta(\mathbf{ h})-U_\zeta(0)\ge \frac{1}{2 }\mathbf{ h}^\intercal{\mathbf{Q}}  \mathbf{ h}
\end{align*}
and in consequence,\vspace{-0.2cm}
\begin{align}\label{eq:Taylor}
 \exp\{ U_\zeta(0)\} \psi(\mathbf{h}) \exp\{-U_\zeta(\mathbf{ h})\} \le \mathcal{P}(\mathbf{ h})\exp\{ -\frac{1}{2 }\mathbf{ h}^\intercal{\mathbf{Q}}  \mathbf{ h}\}.
\end{align}
Because ${\mathbf{Q}}$ is positive definite, the  right-hand side term of  inequality in~\eqref{eq:Taylor} is an integrable majorant of the left-hand term of the inequality, whatever the degree of the polynomial $\mathcal{P}$. Using the point-wise convergence~\eqref{eq:point-wiseCV},  the  Lebesgue's dominated convergence theorem    yields  \vspace{-0.2cm}
\begin{align*}
\lim_{\zeta \to 0}\int  \exp\{ U_\zeta(0)\}\psi(\mathbf{h})\exp\{-U_\zeta(\mathbf{ h})\} d  \mathbf{ h}&=\int \lim_{\zeta \to 0} \exp\{ U_\zeta(0)\}\psi(\mathbf{h}) \exp\{-U_\zeta(\mathbf{ h})\} d  \mathbf{ h}\\
&=\int\psi(\mathbf{h}){\exp\{ -\frac{1}{2}\mathbf{h}^\intercal{\mathbf{H}_{U}} \mathbf{h}} \}d\mathbf{h},
\end{align*}
or equivalently 
\begin{align}\label{eq:cvgen}
\lim_{\zeta \to 0}\int  \psi(\mathbf{h})\exp\{-U_\zeta(\mathbf{ h})\} Z_\zeta^{-1}d  \mathbf{ h}&=\int\psi(\mathbf{h}){ \frac{\exp^{ -\frac{1}{2}\mathbf{h}^\intercal{\mathbf{H}_{U}} \mathbf{h}} }{(2\pi)^{n/2}|\det {\mathbf{H}_{U}}|^{-1/2}}}d\mathbf{h}.
\end{align}
which proves that for $\boldsymbol{\cal H}=\zeta^{-1/2}(\boldsymbol{\tilde\ctheta} - \boldsymbol{ \theta}_{MAP})$, 
\begin{align}\label{eq:convLoiInter}
\boldsymbol{\cal H}  \overset{d}{\to} \mathcal{N}( 0,{\mathbf{H}_{U}^{-1}}).
\end{align}
Moreover, it follows from ~\eqref{eq:changevar2} and the definition of $\boldsymbol{\cal H} $ that
 \begin{align}
\boldsymbol{ \ctheta}=\zeta^{-1/2}(\boldsymbol{\theta}_{MAP}-\boldsymbol{\hat \theta})+\boldsymbol{\cal H} +\boldsymbol{\hat \theta},
\end{align} 
such that to show the sought result it remains to prove that $\zeta^{-1/2}(\boldsymbol{\theta}_{MAP}-\boldsymbol{\hat \theta})$ converges  to zero. Using the definition of the estimate $\boldsymbol{\hat \theta}=\int \boldsymbol{\tilde \theta}\pi_\zeta(d\boldsymbol{\tilde \theta})$, it follows from \eqref{eq:changeVar} that 
$$\zeta^{-1/2}(\boldsymbol{\theta}_{MAP}-\boldsymbol{\hat \theta})=\int  \mathbf{h}\exp\{-U_\zeta(\mathbf{ h})\} Z_\zeta^{-1}d  \mathbf{ h},$$
 and using \eqref{eq:cvgen} we obtain that 
$
\lim_{\zeta \to 0} \zeta^{-1/2}(\boldsymbol{\theta}_{MAP}-\boldsymbol{\hat \theta})=0,
$
which, with \eqref{eq:convLoiInter}, gives the convergence in distribution stated in the proposition. 

Now for the moments. 
We write, for any multi-index $\boldsymbol{p}$,
\begin{align}\label{eq:eqmom1}
E(|\boldsymbol{ \ctheta}|^{\boldsymbol{p}}) = E((|\zeta^{-1/2}(\boldsymbol{ \tilde\ctheta}-\boldsymbol{ \theta}_{MAP})+\boldsymbol{ \theta}_{MAP}|)^{\boldsymbol{p}}) +R,
\end{align}
where the expectation on the right hand side converges to the same moment of $\mathcal{N}( \boldsymbol{ \theta}_{MAP},{\mathbf{H}_{U}^{-1}})$ by \eqref{eq:cvgen}. The remaining term $R$ in \eqref{eq:eqmom1} is a finite sum of terms of the form \vspace{-0.2cm}
$$
C \; E((|\zeta^{-1/2}(\boldsymbol{ \tilde\ctheta}-\boldsymbol{ \theta}_{MAP})+\boldsymbol{ \theta}_{MAP}|)^{\boldsymbol{p_1}}) \,|\zeta^{-1/2}(\boldsymbol{\theta}_{MAP}-\boldsymbol{\hat \theta})- \boldsymbol{ \theta}_{MAP} +\boldsymbol{\hat \theta}|^{\boldsymbol{p_2}},
$$
where the  exectation converges by the previous discussion, and the last factor converges to $0$. $\quad\quad\quad\quad\quad\quad\quad\quad\quad\quad\quad\quad\quad\quad\quad\quad\quad\quad\quad\quad\quad\quad\quad\quad\quad\quad\quad\quad\quad\quad\quad \square$\\

}

 We may immediately deduce the following lemma.\\

\begin{lemma}\label{prop:4}
Assume $\mu$ is Gaussian of mean  $\boldsymbol{\hat \theta}$,  then
$\mu=\mu_\zeta $.\\
\end{lemma}

Under a Gaussian  assumption, the chilled expected  error defined as
$${\mathcal{E}}_{\mu_\zeta,\boldsymbol{\hat \theta}}(\psi)= \int \|\psi(\boldsymbol{ \theta})-\psi( \boldsymbol{\hat \theta} )\|_2 \,\mu_{\zeta}({d\boldsymbol{\theta}}),$$
 is  equal to the  sought one.
The heuristic pursued in this work is to exploit the family of  distributions \eqref{eq:mu0} and \eqref{eq:paramDistrib} to approximate  the expected  error ${\mathcal{E}}_{\mu,\boldsymbol{\hat \theta}}(\psi)$.
Apart from this assumption, the quantity of interest can obviously be rewritten as an expectation with respect to the  measure \eqref{eq:paramDistrib}, choosing the latter as the biasing distribution\vspace{-0.2cm}
 \begin{align}\label{eq:histMethd}
{\mathcal{E}}_{\mu,\boldsymbol{\hat \theta}}(\psi)&\propto \int \|\psi(\boldsymbol{ \theta})-\psi( \boldsymbol{\hat \theta} )\|_2 \,{\exp^{({U_\zeta(\boldsymbol{\hat \theta}+\zeta^{1/2}(\boldsymbol{\theta}-\boldsymbol{\hat \theta})) }- U(\boldsymbol{ \theta} ))}\mu_{\zeta}({d\boldsymbol{\theta}})},
\end{align}
where the symbol $\propto$  denotes an equality up to a multiplicative factor.  It follows that an estimation relying on the samples obtained by a simulation at low-temperature can in principle be used to estimate the sought expected  error. However, the  evaluation of the integral~\eqref{eq:histMethd}  known as the histogram method~\cite{rickman1991temperature},  or of its truncated Taylor  series known as the cumulant method~\cite{phillpot1992temperature}, are unfortunately both inaccurate in terms of estimation variance, making them irrelevant approaches for large temperature differences.

Nevertheless,  the chilled distribution tends to concentrate at $\boldsymbol{\hat \theta}$ as the temperature decreases. It  follows that  ${\mathcal{E}}_{\mu_\zeta,\boldsymbol{\hat \theta}}(\psi)$ is  in some sense a  local approximation of  ${\mathcal{E}}_{\mu,\boldsymbol{\hat \theta}}(\psi)$,
which will be relevant if the posterior is nearly Gaussian. 
Interestingly, the relevance of the local chilled approximation seems to go beyond this situation. Indeed, to the extent of our numerical experiments, the computation of  ${\mathcal{E}}_{\mu_\zeta,\boldsymbol{\hat \theta}}(\psi)$ and of the target quantity ${\mathcal{E}}_{\mu,\boldsymbol{\hat \theta}}(\psi)$ will yield equivalent results. 

Furthermore, as the temperature decreases, our experiments (Section~\ref{sec:numEval}) reveal that there is a significant gain in the convergence speed of the MCMC algorithms, which are used to approximate these expected  errors, and that we present in the next section.  \vspace{-0.2cm}
 
 \section{Chilled Sampling}\label{sec:MCMCgrad}

 The accuracy of a   Monte-Carlo approximation ${\mathcal{E}}_{{\hat \mu_\zeta},\boldsymbol{\hat \theta}}(\psi)$  of ${\mathcal{E}}_{\mu_\zeta,\boldsymbol{\hat \theta}}(\psi)$  will rely on the ability to represent efficiently  the (non-rescaled) chilled  distribution ${\pi_\zeta}$ with an empirical measure,  in our high dimensional context. Traditional sampling approaches  where the normalization constant of ${\pi_\zeta}$ is unknown take the form of Metropolis-Hastings MCMC algorithms.\vspace{-0.cm} 
 
   \subsection{General Metropolis-Hastings algorithms}
Metropolis-Hastings Algorithms rely on a Markov chain defined by a proposal kernel $\mathcal{K}( \boldsymbol{\tilde \theta},d \boldsymbol{\tilde \theta}')$ and a related acceptance probability. Defining the couple distribution 
$
\nu(d \boldsymbol{\tilde \theta},d \boldsymbol{\tilde \theta}')=\pi_\zeta(d \boldsymbol{\tilde \theta})\mathcal{K}( \boldsymbol{\tilde \theta},d \boldsymbol{\tilde \theta}'),
$
the acceptance probability of a state change from $\boldsymbol{\tilde \theta}$ to the proposal $\boldsymbol{\tilde \theta}'$ is defined by\vspace{-0.2cm}
\begin{align}\label{eq:acceptP}
a(\boldsymbol{\tilde \theta},\boldsymbol{\tilde \theta}')= 1 \wedge \frac{\nu(\boldsymbol{\tilde \theta}',\boldsymbol{\tilde \theta})}{\nu(\boldsymbol{\tilde \theta},\boldsymbol{\tilde \theta}')},
\end{align}
where $\nu(d\boldsymbol{\tilde \theta}',d\boldsymbol{\tilde \theta})$ is assumed to be absolutely continuous with respect to $\nu(d\boldsymbol{\tilde \theta},d\boldsymbol{\tilde \theta}')$ and where the minimum of two reals $a$, $b$ is denoted by $a \wedge b$. 
 According to the theory of Metropolis-Hastings~\cite{metropolis1953equation,hastings1970monte},  the samples of the Markov chain $\{{\boldsymbol{\tilde \theta}^i}\}_{i=1}^N$ generated by the   proposal  $\mathcal{K}( \boldsymbol{\tilde \theta},d \boldsymbol{\tilde \theta}')$ with the acceptance probability $a(\boldsymbol{\tilde \theta},\boldsymbol{\tilde \theta}')$  can be used under some additional mixing conditions,  to compute the expectation \vspace{-0.2cm}
\begin{align}\label{eq:MCexpectation}
\int \psi(\boldsymbol{\tilde \theta}) { \pi_\zeta(d\boldsymbol{\tilde \theta})} =\lim_{N\to \infty}\frac{1}{N}\sum_{i=1}^N \psi({{\boldsymbol{\tilde \theta}^i}}).
\end{align}
 For finite $N$, the algorithm yields an approximation, and  in particular the chilled
 expected  error  is approximated\footnote{Memory capacity is  not always sufficient for the storage of  the $N$ samples. In such situations, one may resort to Jensen inequality for an online computation of an upper bound:  
 $${\mathcal{E}}_{{\hat \mu_\zeta},\boldsymbol{\hat \theta}}(\psi) \le \frac{1}{N\sqrt{\zeta}}  \left(N \sum_{i=1}^N \psi({\boldsymbol{\tilde \theta}}^i)^2- \left( \sum_{i=1}^N \psi(\boldsymbol{\tilde \theta}^i) \right)^2\right)^{1/2}.$$ 
 } by\vspace{-0.2cm}
\begin{align}\label{eq:approxMCMC}
{\mathcal{E}}_{{\hat \mu_\zeta},\boldsymbol{\hat \theta}}(\psi)=&\frac{1}{N\sqrt{\zeta}} \sum_{i=1}^N \|\psi({\boldsymbol{\tilde \theta}}^i)-\psi(\boldsymbol{\hat \theta}) \|_2\quad\textrm{with}\quad
\boldsymbol{\hat \theta} =\frac{1}{N}\sum_{i=1}^N {\boldsymbol{\tilde \theta}}^i, 
\end{align}
 in agreement with the change of variable \eqref{eq:changevar2}.
Adjusted random walks  are instances of the class of  Metropolis-Hastings algorithm. They rely on a proposal $\mathcal{K}( \boldsymbol{\tilde \theta},d \boldsymbol{\tilde \theta}')$ independent of the structure of the target posterior, typically a Gaussian proposal
$
\mathcal{K}( \boldsymbol{\tilde \theta},d\boldsymbol{\tilde \theta}')=\mathcal{N}(\boldsymbol{\tilde \theta},\delta t \boldsymbol{\Sigma}_{RW}),
$
where  $\mathcal{N}(\boldsymbol{\tilde \theta},\delta t \boldsymbol{\Sigma}_{RW})$ denotes the multivariate normal distribution of mean $ \boldsymbol{\tilde \theta}$ and covariance $\boldsymbol{\Sigma}_{RW} \in \Rr^{n\times n}$. The particular case where $\boldsymbol{\Sigma}_{RW} =I_n$ corresponds to the standard random walk algorithm. 
 However, random walks are characterized by poor performances when the dimension gets very large.

  \subsection{Metropolis--Adjusted Langevin algorithm (MALA)}
 MCMC algorithms taking into account  the posterior have found to be efficient to explore state spaces of very large dimension. 
In these advanced  methods, the proposal kernel  takes advantage of gradient information of the log posterior in a steepest-descent setting~\cite{roberts2004general,girolami2011riemann,pereyra2015survey}.   
  Among them, the MALA algorithm~\cite{roberts1996exponential} relies on proposals drawn according to 
the Langevin stochastic differential equation (SDE)\vspace{-0.2cm}
$$
\frac{d\boldsymbol{\tilde \theta}}{dt}=-\frac{1}{2}\nabla U_\zeta(\boldsymbol{\tilde \theta})+\frac{dW}{dt},
$$  
 where $W$ is a $n$-dimensional Wiener process. 
 Formally, the above SDE preserves the target  $\pi_\zeta$.
 A discretization of the SDE using an Euler  scheme with  time step $\delta t$ will yield the Gaussian proposal kernel\vspace{-0.2cm}
\begin{align}
\mathcal{K}( \boldsymbol{\tilde \theta},d\boldsymbol{\tilde \theta}')=\mathcal{N}(\boldsymbol{\tilde \theta}-\frac{\delta t}{2}\nabla U_\zeta(\boldsymbol{\tilde \theta}),\delta t  I_n).
\end{align}
We remark that there are alternative to this simple discretization scheme. In particular, semi-implicit  schemes have proved to enhance the acceptance probability under mild conditions~\cite{beskos2017geometric}. Since the SDE preserves $\pi_\zeta$, it is tempting to build our MCMC estimator using the samples obtained by simulating the above discrete-time approximation.  However, it is well known that the discrete-time approximation of the SDE  will differ from the targeted discrete-time dynamics:  decreasing $\delta t$ will reduce the bias, but increase the number of steps to reach stationarity and the correlation among the different samples from $\pi_\zeta$ once stationarity has been reached. A Metropolis-Hastings correction removes the bias. This correction adds however some computational load, in the sense that the function value $U_\zeta$ and its gradient $\nabla U_\zeta$ have to be evaluated for the proposed state $\boldsymbol{\tilde \theta}' $ in addition to the current one $\boldsymbol{\tilde \theta}$ in order to compute the acceptance probability. Moreover high-dimensional problems require in general small $\delta t$. 

The use of  a well-chosen preconditioner is known to be mandatory for  the sampler to remain effective in high dimension~\cite{beskos2017geometric}.  This crucial choice will be examined in the context of our specific meteorological problem in Section~\ref{sec:fBmPre}. Consider a generic preconditioning matrix  $\boldsymbol{\Sigma}_{H} \in \Rr^{n\times n}$. The preconditioned  Langevin SDE is\vspace{-0.2cm}
$$
\frac{d\boldsymbol{\tilde \theta}}{dt}=-\frac{1}{2}\boldsymbol{\Sigma}_{H}\nabla U_\zeta(\boldsymbol{\tilde \theta})+\boldsymbol{\Sigma}_{H}^{\frac{1}{2}}\frac{dW}{dt},
$$  
yielding  the proposal\vspace{-0.2cm}
\begin{align}\label{eq:kernelMC}
\mathcal{K}( \boldsymbol{\tilde \theta},d\boldsymbol{\tilde \theta}')=\mathcal{N}(\boldsymbol{\tilde \theta}-\frac{\delta t}{2}\boldsymbol{\Sigma}_{H}\nabla U_\zeta(\boldsymbol{\tilde \theta}),\delta t \boldsymbol{\Sigma}_{H}),
\end{align} 
using a Euler's scheme time discretization.
The  MALA algorithm iterates between the computing of a proposal step using the kernel~\eqref{eq:kernelMC}, and an accept/reject step according to the acceptance probability \eqref{eq:acceptP} particularized to\vspace{-0.2cm}
\begin{align}\label{probaAccpet}
 \frac{\nu(\boldsymbol{\tilde \theta}',\boldsymbol{\tilde \theta})}{\nu(\boldsymbol{\tilde \theta},\boldsymbol{\tilde \theta}')}&=\exp\{- U_\zeta(\boldsymbol{\tilde \theta})+ U_\zeta(\boldsymbol{\tilde \theta}' )+\frac{1}{2} {\tilde w}^\intercal\boldsymbol{\Sigma}_{H}^{-1} {\tilde w}-\frac{1}{2}{w}^\intercal\boldsymbol{\Sigma}_{H}^{-1}{w}\},
\end{align}
with ${w}=\boldsymbol{\tilde \theta}'-\boldsymbol{\tilde \theta}+\frac{\delta t}{2}\boldsymbol{\Sigma}_{H}\nabla U_\zeta(\boldsymbol{\tilde \theta})$ and ${\tilde w}={w}-\frac{\delta t}{2}\boldsymbol{\Sigma}_{H}\left( \nabla U_\zeta(\boldsymbol{\tilde \theta})+ \nabla U_\zeta(\boldsymbol{\tilde \theta}' ) \right)$.
As exposed in the next section, this algorithm is in fact a particular instance of a more general and powerful class of algorithms based on Hamiltonian dynamics. 

 \subsection{Hamiltonian Monte Carlo algorithm (HMC)}\label{sec:HMC}
HMC is a  powerful state-of-the-art algorithm as it enjoys desirable scaling properties for high-dimensional  problems \cite{neal2011mcmc}. 
To define HMC, we need to introduce the Hamiltonian functions that can be written as follows:\vspace{-0.2cm}
$$
(\boldsymbol{\tilde \theta},\boldsymbol{\xi}) \mapsto U_\zeta(\boldsymbol{\tilde \theta})+ K(\boldsymbol{\xi}),
$$
where  minus the log posterior probability $U_\zeta(\boldsymbol{\tilde \theta})$ is complemented by a function   $ K: \Rr^n\to \Rr$ of some auxiliary  state variable  $\boldsymbol{\xi}\in \Rr^n$ called {\it kinetic energy}. Classically, the exponential of minus this function is up to a normalization constant a centered Gaussian law of covariance $\boldsymbol{\Sigma}_{H}^{-1}$,  so that  the kinetic energy corresponds to  \vspace{-0.2cm}
$$
 K(\boldsymbol{\xi})=\frac{1}{2}\boldsymbol{\xi}^\intercal\boldsymbol{\Sigma}_{H}\boldsymbol{\xi}.
$$
Hamilton's equations can then be written as follows:
\begin{align*}
\frac{d\boldsymbol{\tilde \theta}}{dt}&=\boldsymbol{\Sigma}_{H}\boldsymbol{\xi},\quad 
\frac{d\boldsymbol{\xi}}{dt}=-\nabla U_\zeta(\boldsymbol{\tilde \theta}),
\end{align*}
A leapfrog scheme is known to be relevant  for time-discretization  of these equations as it preserves volume exactly and is time-reversible. It yields the recursion
\begin{align}\label{eq:leapfrog}
\boldsymbol{\tilde \theta}(t+\delta t)&=\boldsymbol{\tilde \theta}(t)-\frac{\delta t^2}{2}\boldsymbol{\Sigma}_{H}\nabla U_\zeta(\boldsymbol{\tilde \theta}(t))+{\delta t} \boldsymbol{\Sigma}_{H} {\boldsymbol{\xi}(t)},\\
\boldsymbol{\xi}(t+\delta t)&=\boldsymbol{\xi}(t)-\frac{{\delta t}}{2} \left( \nabla U_\zeta(\boldsymbol{\tilde \theta}(t))+\nabla U_\zeta(\boldsymbol{\tilde \theta}(t+\delta t))\right), 
\end{align}
with the initial potential state set to  $\boldsymbol{\tilde \theta}(t_0)=\boldsymbol{\tilde \theta}$ and an initial kinetic state $\boldsymbol{\xi}(t_0)$ drawn according to the Gaussian law
$
\mathcal{N}(0, \boldsymbol{\Sigma}_{H}^{-1} ).
$
After performing $L$ steps with the leapfrog recursion,  the Hamiltonian dynamics  yields the new potential state  $\boldsymbol{\tilde \theta}'=\boldsymbol{\tilde \theta}(t_0+L\delta t)$ and the  kinetic state $\boldsymbol{\xi}'=\boldsymbol{\xi}(t_0+L\delta t)$. 

The HMC algorithm consists in alternating an Hamiltonian dynamics proposal and  a Metropolis accept/reject procedure: a new state  $\boldsymbol{\tilde \theta}'$ is accepted according to the probability~\eqref{eq:acceptP} particularized for  HMC 
\begin{align*}
a(\boldsymbol{\tilde \theta},\boldsymbol{\tilde \theta}')= 1 \wedge \exp\{U_\zeta(\boldsymbol{\tilde \theta})-U_\zeta(\boldsymbol{\tilde \theta}')+K(\boldsymbol{\xi})-K(\boldsymbol{\xi}')\}.
\end{align*}
The HMC algorithm is a generalization of the MALA algorithm, as the latter corresponds to HMC in the particular case where  only one leapfrog recursion is used, that is to say  $L=1$ and we substitute $\delta t$ by $\sqrt{\delta t}$. Indeed, noticing that  $\boldsymbol{\Sigma}_{H}\boldsymbol{\xi}(t_0) \sim  \mathcal{N}(0,\boldsymbol{\Sigma}_{H}),$  it is easy to see that in  this particular setting the proposition kernel $\mathcal{K}( \boldsymbol{\tilde \theta},d\boldsymbol{\tilde \theta}')$ is the same as the one in MALA \eqref{eq:kernelMC}, so as the acceptance probability \eqref{probaAccpet}.

{Note that in MALA, due to the structure of the Metropolis ratio, the gradient  $\nabla U_\zeta $ in the proposal can be replaced \textit{mutatis mutandis} by any vector field $T$ without modifying the target distribution. In HMC, however, the vector field in the proposal must derive from a gradient $T= \nabla \tilde{U}$ for the leapfrog scheme to be symplectic and preserve volume. In the appendix, we propose a fast gradient computation which involves  a very small approximation based on finite differences. In theory, this could lead to a slight bias in the HMC method, but we did not observe any significant difference from the MALA method.}

Finally, we point out that the numerical simulations (see Section~\ref{sec:numEval}) show that for a given acceptance rate, a temperature tending towards zero does not have much impact on the (relative) time step of the MALA or HMC simulation, while it significantly accelerates the convergence of the algorithm. The acceleration in convergence speed is caused exclusively by gradient nonlinearities in the posterior log-density, as temperature does not affect the approximation of the expected  error.
Because of the invariance of the time step with temperature, we are led to believe that the acceleration observed by chilling is exclusively related to the slow direction (large variance) of the posterior distribution.

%
%
%

 \subsection{The proposed algorithm}

The complete methodology for uncertainty quantification by chilled  sampling proposed in this work is taken up by the algorithm presented below. We restrict ourselves to the best-performing gradient-based sampler, the HMC algorithm (see Section~\ref{sec:HMC}). \medskip

{\bf Algorithm 1 (Uncertainty quantification)}
\begin{algorithmic}[1]
 \Require point estimate $\boldsymbol{\hat \theta}$, low temperature $\zeta \ll 1$, preconditioning parameter $H$ 
\State Simulate by HMC   the  (non-rescaled) variable $\boldsymbol{\tilde \ctheta} $ at temperature  $\zeta$ defined in   \eqref{eq.pizeta}

\State {Rescale variable  $\boldsymbol{\tilde \ctheta} $ into $\boldsymbol{ \ctheta}$ using \eqref{eq:changevar2}}

\State{Compute the chilled expected error ${\mathcal{E}}_{{\hat \mu_\zeta},\boldsymbol{\hat \theta}}(\psi)$ defined in \eqref{eq:approxMCMC}}  
\end{algorithmic}

\section{The Meteorological Inverse Problem of Interest}\label{Sec:Model}

\subsection{Bayesian model for AMVs}
We describe in the following the Bayesian model used for the estimation of AMVs, given a stack of couples of noisy images, the images being defined  partially on the pixel grid. The strategy is to estimate jointly a displacement field and the couples of images on the entire pixel grid from the noisy and partial  observations.
\subsubsection{Unknowns}
The  variables of interest are a pair of stack of images $(\xx_{t_0}^\star,\xx_{t_1}^\star)\in (\Rr^{km})^2$ and the AMVs, \ie, a displacement field $\mathbf{d}^\star\in \Rr^{2m}$. The  stacks $\xx_{t_0}^\star$ and $\xx_{t_1}^\star$  are composed of $k$-variate fields in $\Rr^m$. For $\ell=1,\ldots,k$, the $\ell$-th  field gathered in $\xx_{t_0}^\star$ and $\xx_{t_1}^\star$ will be denoted by $\xx_{t_0}^{\ell,\star}$ and $\xx_{t_1}^{\ell,\star}$. The components of the vectors $\xx_{t_0}^{\ell,\star}$ and $\xx_{t_1}^{\ell,\star}$ are associated to spatial coordinates on the pixel grid  $$\Omega_m=\{q\in \Omega : q=\varkappa(s), s=1,\ldots,m\},$$
where the bi-dimensional domain is $\Omega$ and  $\varkappa(s)$ is the function returning the spatial position corresponding to index $s$. {As a consequence, $\xx_{t}^{\ell,\star}(s)$ will denote the $\ell$-th field at position $\varkappa(s)$ and time $t$.}

A standard assumption is to consider that each image pair $(\xx_{t_0}^{\ell,\star},\xx_{t_1}^{\ell,\star})$ {represents} the continuous solution $\tvt(q,t) \in  \mathcal{C}^1(\Omega \times \mathbb{R})$ taken on the points of the grid $\Omega_m$, at times $t_0$ and $t_1$ (with $t_0 < t_1$) of the transport  equation of  initial condition $\tvt_{t_0}(q)$, so-called in the image processing literature ``optic-flow equation''
\begin{equation}\label{eq:OFC}
 \left\{\begin{aligned}
&\frac{\partial \tvt}{\partial t}(q,t)  + \wwr(q,t) \cdot\nabla  \tvt(q,t) =0\\
&\tvt(q,t_0)=\tvt_{t_0}(q)
\end{aligned}\right.
\end{equation}
where $\wwr(q,t)$ is the transportation field that verifies $\ww(q,t) \in  \mathcal{C}^1(\Omega \times \mathbb{R})$.
It is well known that under mild conditions, when $dt\triangleq t_1-t_0$ is a small increment, we can write the warping constraint\footnote{
It follows from \eqref{eq:OFC} that $\tvt_{t_0}(q) =\tvt({\bf Q}^{t_1}_{t_0}(q),t_1) $ where the function $t \rightarrow {\bf Q}_{ t_0}^t(q)$, known as the characteristic curves of the partial differential equation \eqref{eq:OFC}, is  the solution of the system~\cite{raviart1983introduction}:
 \begin{equation*}
 \left\{\begin{aligned}
&\frac{d }{d t}{{\bf Q}}_{ t_0}^t(q)  = \wwr({{\bf Q}}_{ t_0}^t(q),t), \\
&{{\bf Q}}_{ t_0}^{t_0}(q)=q.
\end{aligned}\right.
\end{equation*}
 Assuming that  $\int_{t_0}^{t_1}\wwr({{\bf Q}}_{ t_0}^s(q),s)ds=dt \wwr(q,t_0)$,  then  we  obtain by time integration \eqref{eq:warpingDFD}.}
 \begin{align}\label{eq:warpingDFD}
 \tvt_{t_0}(q) =\tvt_{t_1}(q+ dt\, \wwr(q,t_0)).
 \end{align}
In order to build a warping constraint  for each image couple $(\xx_{t_0}^{\ell,\star},\xx_{t_1}^{\ell,\star})$, we will need to assume a continuous model for interpolating the images $\xx_{t_1}^{\ell,\star}$ outside of $\Omega_m$. Taking the constraints~\eqref{eq:warpingDFD} at points in $\Omega_m$  and making the idenification $\mathbf{d}^\star(s)=dt\, \wwr(\varkappa(s),t_0)$, we rewrite  the warping model  as
\begin{align}\label{eq:warpingModel}
\mathbf{x}_{t_0}^\star(s_0)=\mathcal{W}_{s_0}(\xx_{t_1}^\star, \mathbf{d}^\star), \quad s_{0}=1, \ldots, m,
\end{align}
where operator $\mathcal{W}: \Rr^{km} \times \Rr^{2m}\rightarrow \Rr^{km} $  in \eqref{eq:warpingModel} warps the stack of images $\xx^\star_{t_1}$
according to the  displacement  $\mathbf{d}^\star\in \Rr^{2m}$. 
 The $s$-th component output $\mathcal{W}_{s}(\xx_{t_1}^\star,\mathbf{d}^\star): \Rr^{km} \times \Rr^{2m} \to \Rr^k$ of operator $\mathcal{W}$ is the 
 function   defined as  
\begin{align}\label{eq:splineRepr}
\mathcal{W}_{s_0}(\xx_{t_1}^\star,\mathbf{d}^\star) =\sum_{s_1\in \mathcal{V}(\varkappa(s_0)+\mathbf{d}^\star(s_0))} \xx_{t_1}^\star(s_1)\varphi_{s_1}(\varkappa(s_0)+\mathbf{d}^\star(s_0)),
 \end{align}
where $\mathcal{V}(\varkappa(s)+\mathbf{d}^\star(s))$ denotes a subset of  indices corresponding to the ``neighborhood'' of point $\varkappa(s)+\mathbf{d}^\star(s)$. The family $\{\varphi_s\}_{s=1}^m$ with  $\varphi_s: \Omega_m \to \Rr $ may for example be bi-dimensional cubic cardinal   splines  interpolation functions  \cite{Unser91}. 
Note that the image $\xx_{t_0}^\star$ is a deterministic function of $\xx_{t_1}^\star$ and $\mathbf{d}^\star$.  We remark that $\mathcal{W}$ is linear in its first argument and  non-linear in its second one as long as $\varphi_s$'s are  non-linear.  Operator $\mathcal{W}$  can be generalized to take into account   for all the modifications of the field $\mathbf{x}_{t_0}^\star$ which
cannot be inferred from $\xx_{t_1}^\star$, including error on the physical model or interpolation errors~\cite{heas2016efficient}. 

 Let us   remark  that the  transport model~\eqref{eq:warpingModel} constitutes  a relevant model  frequently  encountered  in physics. In fluid mechanics, it describes the non-diffusive advection of a passive scalar by the flow \cite{Liu08}.  In atmospheric sciences, under the assumption of negligible vertical winds and diabatic heating, the transport equation describes the horizontal displacement of pressure-averaged fields of  temperature, specific humidity or ozone~\cite{Holton92}, as detailed in Appendix~\ref{app:modelMeteo}.


\subsubsection{Partial image observations}
The couple $(\mathbf{d}^\star,\xx_{t_1}^\star)$, which we will refer as  the ``ground truth'',  generates noisy and partial observations:  we observe a subset of components of $\xx^\star_{t_0}=\mathcal{W}(\xx^\star_{t_1}, \mathbf{d}^\star)$ and $\xx^\star_{t_1}$ up to an additional noise to be specified later.   Let   $ \Omega^{t_0}_{obs},\,  \Omega^{t_1}_{obs} \subseteq \Omega_m$ denote the set of spatial locations  related to  the observed components of $\xx_{t_0}^\star$ and $\xx_{t_1}^\star$.
The set of observations is  $$\yy=\{\yy_t^{obs}({s})\in \Rr^k : s \in  \Omega^{t}_{obs} , t\in \{t_0,t_1\}\} \in \mathcal{Y},$$
where the $\yy_t^{obs}({s})\in \Rr^k$ is the observed components $\xx^\star_t(s)$ up to some additional noise. We will denote by  $\Omega_{obs}= \Omega^{t_0}_{obs} \cap \Omega^{t_1}_{obs}$ the set of consecutively observed pixels. Note that our observation model assumes that if $s\in \Omega_{obs}^t$, then we observe the $k$ components of $\xx^\star_t(s)$.

\subsubsection{Likelihood model} \label{sec:likeli}

The problem consists in recovering the unknown ground truth vector $\boldsymbol{\theta}^\star=\begin{pmatrix}\mathbf{d}^\star\\\xx_{t_1}^\star\end{pmatrix}$  from the partial observations $\yy$.   Consider the variable  $\boldsymbol{\theta}=\begin{pmatrix}\mathbf{d}\\\xx_{t_1}\end{pmatrix} \in  \Rr^{n}$ with \vspace{-0.2cm}$$n=(2+k)m.$$ The estimation  classically relies on a likelihood function  linking $\boldsymbol{\theta}$ to a random variable $\yy \in \mathcal{Y}$, standing for observations.   
To this aim, we define the residual  function $\boldsymbol{\delta}_{}: \Rr^n \times \mathcal{Y} \to \Rr^{2km}$ such that $\boldsymbol{\delta}_{}(\boldsymbol{\theta},\yy)=\begin{pmatrix}{\boldsymbol{\delta}_{t_0}}(\boldsymbol{\theta},\yy)\\ {\boldsymbol{\delta}_{t_1}}(\boldsymbol{\theta},\yy)\end{pmatrix}$, whose $\iota(t,s)$-th component is  defined for $t=t_0,t_1$, $s\in \Omega_m$ with the indexing $\iota :\{t_0,t_1\} \times \Omega_m \to \mathbb{N}$
 as: \vspace{-0.2cm}
 \begin{align}\label{eq:defDeltax}
{\boldsymbol{\delta}_{\iota(t,s)}}(\boldsymbol{\theta},\yy)=
\left\{\begin{aligned}
&\xx_t({s})-\yy_t^{obs}({s}) \quad \textrm{if}\quad {s}\in \Omega^{t}_{obs},\\
&0  \quad \hspace{2.1cm}\textrm{otherwise}.
 \end{aligned}\right. 
\end{align}

The non-zero components of  the residual are function of $\boldsymbol{\theta}$  and of the observations $\yy$, corrupted by some noise of density~$f$. We  define the negative log-likelihood $\phi: \Rr^n\times \mathcal{Y} \to \Rr$ as:\vspace{-0.3cm}
$$\phi(\boldsymbol{\theta};\yy)=-\log f(\boldsymbol{\delta}(\boldsymbol{\theta},\yy)).$$ 
For example, in the case of  an i.i.d. centered Gaussian noise of variance $\beta^{-1}$,
the  model generating the observations is 
\begin{align*}
\yy_{t_0}^{obs}({s})&=\mathcal{W}_{s}(\xx_{t_1}^\star, \mathbf{d}^\star)+\mathcal{N}(0,\beta^{-1}), \quad  s \in  \Omega^{t_0}_{obs},\\
\yy_{t_1}^{obs}({s})&=\xx^\star_{t_1}(s)+\mathcal{N}(0,\beta^{-1}), \quad  s \in  \Omega^{t_1}_{obs},
\end{align*}
yielding the the negative log-likelihood
\begin{align}\label{eq:likelyGauss}
\phi(\boldsymbol{\theta};\yy)=\beta \|\boldsymbol{\delta}(\boldsymbol{\theta},\yy)\|^2_2.
\end{align}
More involved schemes use alternative noises to take into account non-quadratic deviations, see references in~\cite{Butler:ECCV:2012}. 


A straightforward criterion to estimate the unknown fields  from the observations is to maximize the likelihood: ${\arg \min}_{\boldsymbol{\theta}\in \Rr^n}\,\phi(\boldsymbol{\theta};\yy).$  This  severely ill-conditioned  problem is known as the {\it aperture problem} in the computer vision community \cite{Kadri13}. In the following, we will  classically rely on a Bayesian framework.

\subsubsection{Posterior  model}\label{sec:prior}
Given some prior, the posterior distribution is  defined by \eqref{eq:posterior}.
 Many priors  have been proposed in the  literature for $\boldsymbol{\theta}$, that is to say for displacement fields and image intensity functions.   Gaussian priors $\nu_0$ of negative logarithm of the form $\varphi(\boldsymbol{\theta})= \boldsymbol{\theta}^\intercal\boldsymbol{\Sigma}_0^{-1}\boldsymbol{\theta}$ will be of  interest in our numerical simulations.
 More precisely,  the random variable is $\boldsymbol{\theta}=(\mathbf{d}^\intercal,\xx_{t_1}^\intercal)^\intercal\in \Rr^n$, with $n=(2+k)m$. We assume that $\mathbf{d}\in \Rr^{2m}$ and $\xx_{t_1}\in \Rr^{km}$ are independent, so that the prior covariance decomposes as\vspace{-0.2cm}
  \begin{equation}\label{priorCov}
  \boldsymbol{\Sigma}_0= \begin{pmatrix} \alpha\boldsymbol{\Sigma}_{\mathbf{d}} &\mathbf{0} \\ \mathbf{0} &\gamma \boldsymbol{\Sigma}_{\mathbf{x}} \end{pmatrix},
  \end{equation}
where we have introduced the two parameters $\alpha,\gamma >0$ with $\boldsymbol{\Sigma}_{\mathbf{d}} \in \Rr^{2m \times 2m}$ and   $\boldsymbol{\Sigma}_{\mathbf{x}} \in \Rr^{km \times km}$.
 The covariance matrix $\boldsymbol{\Sigma}_0$  will have  a  structure specific to the statistics of the  field of interest. In particular for fluid flows, standard choices for the covariance matrix correspond to smoothing the gradient of the divergence and vorticity of the flow \cite{Suter94}.  In this work we will focus on more recent  schemes   structuring  long-range interactions of the displacement field using bivariate isotropic fractional Brownian motion (fBm) priors specified by two  hyper-parameters\footnote{Usually, the tuning of the prior hyper-parameters and the log likelihood parameter $\beta$ is done manually, guided by some ground truth data and the expert knowledge of meteorological scientists. However, there are also advanced statistical methods for hyper-parameter inference in optical-flow problems~\cite{krajsek2006maximum,heas2011bayesian,heas2012bayesian,stoll2017time}. In this work, we will assume that the hyper-parameters are known, as the topic of their inference is outside the scope of this paper.}~\cite{Tafti11,Heas14}. \medskip

 The following remark points out that the proposed chilling strategy manages naturally  the traditional optic-flow setting where the likelihood and the prior hyper-parameters are only known up to a multiplicative factor.\\
  
\begin{remark}\label{rem:3}
The computation of an estimate of the MAP \eqref{objFunction5pre}  computed by optic-flow algorithms does not need the explicit knowledge of the  three parameters $\alpha$, $\beta$ and  $\gamma$. The definition of the MAP only needs the determination of the  ratios $\alpha'=\alpha/\beta$ and $\gamma'=\gamma/\beta$:
\begin{align}\label{eq:typicOF}
& \underset{ \boldsymbol{\theta}\in \Rr^n}{\arg \min} \{\phi(\boldsymbol{\theta})+\varphi(\boldsymbol{\theta}) \}. \nonumber \\
&=\underset{ \boldsymbol{\theta}\in \Rr^n}{\arg \min} \{\|\boldsymbol{\delta}(\boldsymbol{\theta},\yy)\|^2_2 + \boldsymbol{\theta}^\intercal\begin{pmatrix} {\alpha'}\boldsymbol{\Sigma}_{\mathbf{d}}^{-1} &\mathbf{0} \\ \mathbf{0} &{\gamma'} \boldsymbol{\Sigma}_{\mathbf{x}} ^{-1}\end{pmatrix}\boldsymbol{\theta} \}
\end{align}
Therefore, the proposed chilled MC simulation remains valid in the case  where parameter $\beta$ is unknown (which is often the case in practice) with the tuning of the temperature $1/\zeta$  substituted by the tuning of the ratio ${\beta}/{\zeta}$.\\
\end{remark}

\subsection{fBm preconditioning}\label{sec:fBmPre}
In addition to chilling, another idea to circumvent the curse of dimensionality  is to structure and adjust the preconditioner $\boldsymbol{\Sigma}_{H}$ according to the posterior  and the  observable $\psi$. For an observable being the identity function,
the posterior covariance $\boldsymbol{\Sigma}_{\pi_\zeta}=\int (\boldsymbol{\tilde \theta}-\boldsymbol{\hat \theta})(\boldsymbol{\tilde \theta}-\boldsymbol{\hat \theta})^\intercal\pi_\zeta(d\boldsymbol{\tilde \theta}) ,$  constitutes in principle  an ideal choice of preconditioner for gradient-based MCMC samplers~\cite{beskos2009mcmc,cotter2013mcmc}. Note however that choosing this ideal preconditioner might not always be optimal in terms of speed of convergence of   \eqref{eq:approxMCMC} towards the  expectation ${\mathcal{E}}_{\mu_\zeta,\boldsymbol{\hat \theta}}(\psi)$. Indeed, the accuracy of  the approximation \eqref{eq:approxMCMC} depends not only on the preconditioner but also on the chosen observable $\psi$. 

 Using the posterior covariance for preconditioning presents some difficulties. First the posterior covariance is most often not available in closed-form, and one can only rely on approximations based on the past samples of the MCMC algorithms.  Next, the gradient-based MCMC algorithms evaluate the inverse of the preconditioner, \ie, invert  the posterior covariance matrix approximation, which could require a prohibitive complexity $\mathcal{O}(n^3)$. Finally, one should also be able to sample a Gaussian with the posterior covariance matrix approximation.   Some recent works face these difficulties by proposing ``on the fly'' low-rank  approximation updates and diagonal inverse approximations of the target covariance $\boldsymbol{\Sigma}_{\pi_\zeta}$ with a complexity  significantly lower than cubic~\cite{luan2020langevin}. The proposed approach results in significant computational and memory  overload, which can be critical in high-dimensional configurations.   Another route that has attracted interest is to substitute  the ideal preconditioner by the prior covariance matrix $\boldsymbol{\Sigma}_0$~\cite{cotter2013mcmc}. This strategy is particularly advantageous in the case of closed-form Gaussian priors.  

{As an alternative}, we propose  to pick the preconditioner in the family of   matrices  corresponding to covariances of  isotropic bi-dimensional fBms of Hurst exponent $H \in \Rr_+$, truncated on the pixel grid. The definition of these fBms in terms of Fourier or wavelet series can be found in~\cite{Tafti11,Heas14}. Their tractable covariance structure is  provided  in  Appendix~\ref{app:fBm}. Their covariance is parametrized by a single parameter\footnote{ Up to a multiplicative constant which we set here to a unit value. Indeed this factor has no impact on the algorithms as the preconditioner is systematically multiplied by the MCMC discretization time step $\delta t$, which is a free parameter of the algorithms.}, namely  exponent $H$.
This heuristic choice is shown to be experimentally relevant as far our numerical simulations are concerned, see Section~\ref{sec:numEval}.
The idea behind this choice is that there hopefully exists an fBm instance whose covariance matches in some sense  the eigenvectors with smallest variance of the posterior covariance, and is  better suited than the prior covariance. 
Thanks to the existence of  fast algorithms, computing the product of the fBm covariance or its inverse with  a vector, 
or sampling the fBm, are tractable operations performed in  $\mathcal{O}(m \log m)$.  
In consequence, the use of such conditioners generates no significative overload compared to state-of-the art optic-flow algorithms, which rely on a gradient descent exhibiting a  linear complexity (see references in~\cite{Butler:ECCV:2012}), or log linear (as in~\cite{Heas14}).\medskip

\subsection{Optimization issues}

The computation of the Laplace approximation or the initialization of  the gradient-based MCMC algorithms both need to estimate the MAP. An estimate of the MAP is accessed by solving the minimization problem~\eqref{eq:typicOF}, which constitutes a typical optic-flow estimation problem. We use a standard limited-memory quasi-Newton optimization scheme~\cite{wright1999numerical} to solve this differentiable unconstrained minimization problem\footnote{For more involved situations where the log of the posterior is non-differentiable or defined over a restriction of $\Theta$, relevant local minima are computed using modern constrained optimization techniques~\cite{Bertsekas99}.}. As proposed in~\cite{derian2013wavelets}, in order to estimate accurately large displacements,  the optimization procedure avoids the heuristic multiresolution optic-flow initialization, and relies instead on wavelet  expansions (Coiflets with 10 vanishing moments) of the displacement variable $\mathbf{d}$. An analogous expansion is used to expand the   image variable $\xx_{t_1}$ accordingly. Appendix~\ref{sec:gradient}  details fast evaluation procedures  computing the  gradient of the log posterior,  similar to the one proposed in~\cite{Heas14}. The fast evaluation  relies on the fast wavelet transform (FWT) and on the fast Fourier transform (FFT) of complexity $\mathcal{O}(m \log m)$.

Particularized to the above posterior model,
Appendix~\ref{rem:CondIndLaplace} details the  computation of the EVD of the Hessian of the log posterior, needed by the Laplace approximation presented in Section~\ref{sec:laplace}.

In addition to fast gradient evaluation, the MCMC algorithms presented in Section~\ref{sec:MCMCgrad} need
 efficient sampling procedures adapted to the proposed fBm preconditioning. 
FBm sampling   is performed in agreement with the wavelet representation detailed in the appendix, using a method analogous to the one proposed in~\cite{Heas14}:  wavelet coefficients are sampled according to standard Gaussian white noise, the fBms realizations are then synthesized by first, application of an inverse FWT to the  coefficient vector, then by application of an FFT followed by a fractional differentiation in Fourier domain and finally by application of an inverse FFT.  For the random walk where $\boldsymbol{\Sigma}_{RW} =\boldsymbol{\Sigma}_{H} $, or for preconditioning the  MALA and the HMC simulation,  the matrix-vector product with the fBm covariance matrix  is computed in an analogous manner in the Fourier domain.

\section{Numerical Simulations}\label{sec:numEval}
The following sections describe the numerical framework used for evaluation in the context of  synthetic and real scenarios.  

\subsection{Evaluation criteria}\label{sec:evalCriteria}

 A common criterion  to compare  the accuracy of optic-flow   estimates  is the average endpoint error~\cite{Butler:ECCV:2012}.  These end-point errors are defined as  $\|\psi(\boldsymbol{\theta^\star})-  \psi( \boldsymbol{\hat \theta} )\|_2$ for $\psi$ in the set $\mathcal{P}(\Omega_m)$ of functions gathering the bi-variate components of the displacement field $\mathbf{d}$, \ie,
\begin{equation}\label{eq:mappingPixels}
\mathcal{P}(\Omega_m)=\{{\psi} :\psi(\boldsymbol{\theta})=\mathbf{ d}(j) \in \Rr^2, {j: \varkappa(j) \in\Omega_m}\}.
\end{equation}
So, to calculate the expected error \eqref{eq:varianceExpectedError}, or its chilled or Laplace approximation, with respect to the displacement field $\mathbf{d}$, the $\psi$ will be chosen in this set.  
For instance, the chilled expected error ${\mathcal{E}}_{{\mu_\zeta},\boldsymbol{\hat \theta}}(\mathbf{d}(j))$ for the bi-variate vector $\mathbf{d}(j)$ with $j: \varkappa(j) \in\Omega_m$ corresponds to the test function $\psi(\boldsymbol{\theta})=\mathbf{ d}(j)$.
 \medskip
On the other hand, since we're jointly estimating image intensities, other interesting $\psi$ will be functions returning the $k$-th components of the image intensity stack, taken at the points of the image grid $\Omega_m$.

 In our experiments, we  evaluate the weighted average endpoint error $$\textrm{EPE}(\Omega,{w}_{\hat \mu, \boldsymbol{\hat \theta} }^{p},\boldsymbol{\theta^\star}-  \boldsymbol{\hat \theta})$$ defined in \eqref{eq:MSE},  where  $\boldsymbol{\hat \theta}$ is the estimate (given by the Laplace or the MCMC approximation), and where    the  weighting function ${w}_{\hat \mu, \boldsymbol{\hat \theta} }^{p}$ defined in \eqref{eq:weightL1_} - \eqref{eq:weightL0_} is computed  for an expected error related to a posterior approximation $\hat \mu$. The latter is either  the Gaussian approximation ${{\mu_0}}$ (given by the Laplace approximation) or  the empirical measure ${{\hat \mu_\zeta}}$ for  temperatures $\zeta>0$ (given by the MCMC approximation). 
With regard to the choice of the spatial domain $\Omega$,  we  focus our interest either on  the full pixel grid $\Omega_m$, or on the sub-domain $\Omega_{obs}$, \ie,  the sub-set of pixels observed at consecutive times. 
In summary, we will consider the following  criteria:
\begin{itemize}
\item  the   {standard EPE} defined as ${\textrm{EPE}}(\Omega_m, \mathbf{1},\boldsymbol{\theta^\star}-  \boldsymbol{\hat \theta}) $,
\item  the   {$p$-weighted EPE} defined as ${\textrm{EPE}}(\Omega_m,{w}_{\hat \mu, \boldsymbol{\hat \theta} }^{p},\boldsymbol{\theta^\star}-  \boldsymbol{\hat \theta}) $ for $p=1,2,$
\item the   {masked EPE} defined as ${\textrm{EPE}}(\Omega_{obs}, \mathbf{1},\boldsymbol{\theta^\star}-  \boldsymbol{\hat \theta}) $,
\item  the   {sparse EPE} defined as ${\textrm{EPE}}(\Omega_m,{w}_{\hat \mu, \boldsymbol{\hat \theta} }^{0},\boldsymbol{\theta^\star}-  \boldsymbol{\hat \theta}) $ with $\tau={\sharp\mathcal{P}(\Omega_{obs})} $
\item  the  {sparse masked EPE} defined as ${\textrm{EPE}}(\Omega_{obs}, {w}_{\hat \mu, \boldsymbol{\hat \theta} }^{0},\boldsymbol{\theta^\star}-  \boldsymbol{\hat \theta}) $  with $\tau={\sharp\mathcal{P}(\Omega_{obs})} /2$.\\
\end{itemize}
\medskip
 Their calculation, defined in Section~\ref{sec:eQerror}, is taken up below.\medskip

{\bf Algorithm 2 (Weighted endpoint error)}
\begin{algorithmic}[1]
 \Require domain $\Omega$, expected error ${\mathcal{E}}_{{\hat \mu},\boldsymbol{\hat \theta}}(\psi)$,  truth $\boldsymbol{\theta^\star}$, point estimate $\boldsymbol{\hat \theta}$, order $p$
\State  Compute the weights ${w}_{\hat \mu,\boldsymbol{\hat \theta}}^{p}(\psi)$ defined in \eqref{eq:weightL1_} and \eqref{eq:weightL0_} using  ${\mathcal{E}}_{{\hat \mu},\boldsymbol{\hat \theta}}(\psi)$ and related   $c_p$ (see Section~\ref{sec:eQerror})  
\State {Compute $\textrm{EPE}(\Omega,{{w}_{\hat \mu,\boldsymbol{\hat \theta}}^{p}(\psi)},\boldsymbol{\theta^\star}-  \boldsymbol{\hat \theta})$ defined in \eqref{eq:MSE}}
\end{algorithmic}

 Let us make a few comments on these criteria.
The most important remark is that, unlike the standard or masked EPE, the other criteria assess the accuracy of the error estimate, in addition to the accuracy of the posterior mean estimate. Next, the standard or $p$-weighted EPE are global criteria, whereas the other criteria evaluate accuracies on a subdomain of the pixel grid. Moreover, several of these criteria can be compared to each other. Indeed, because  constant unit weights satisfy  the constraints  \eqref{eq:const0} and  \eqref{eq:const1}, we remark that the expectation of the standard EPE is always greater than the expectation of the $p$-weighted EPE. Similarly, because  constant unit weights on the subdomain $\Omega_{obs}$ satisfy  the constraints  \eqref{eq:const2},  the expectation of the masked EPE is always greater than the expectation of the sparse EPE. As a last remark, it is clear that the expectation of the sparse masked EPE is lower than the expectation of the masked EPE, and the latter is likely to be lower than the expectation of the standard EPE itself, if the unobserved pixels are related to the larger errors.

\subsection{Data benchmarks}
We detail hereafter different  numerical settings,  ranging from simulated data-sets  to real-world meteorological observations. We use a pixel grid $\Omega_m$ of dimension $m=2^7 \times 2^7$, $k=3,\, n=2^{18}$ and the  following  benchmarks:
\begin{itemize}
\item  {\bf Experiment \#1:}  Synthetic (and incomplete) observations  of a passive scalar field advected by a simulated turbulent motion   ;
\item {\bf Experiment \#2:}  Real-world observations of pressure-averaged atmospheric  humidity, temperature and ozone concentration   at medium altitude (average of the quantities between the isobaric levels of 600 and 700 hPa). Observations are provided,  together with the corresponding ground truth horizontal motion field, by the operational numerical model of the European Centre for Medium-Range Weather Forecasts (ECMWF)~\cite{temperton2001two};
\item {\bf Experiment \#3:}  Real-world (and incomplete) observations  of pressure-averaged atmospheric humidity, temperature and ozone concentration  at medium altitude. The observations are provided by the Infrared Atmospheric Sounding Interferometer (IASI) of Metop-A and Metop-B satellites~\cite{borde2019winds}, while the ground truth is assumed to be the synchronized ECMWF numerical model. \vspace{-0.2cm}\\
\end{itemize}


 For experiment \#1,   the target distribution is the posterior \eqref{eq:posterior} constructed using  the likelihood~\eqref{eq:likelyGauss}  and  the  prior  \eqref{priorCov}  of parameter $\alpha$ and $\gamma$. Here    $\gamma \boldsymbol{\Sigma}_x$ is diagonal constant, while $\alpha \boldsymbol{\Sigma}_{\mathbf{d}}$ corresponds to the covariance of a Gaussian prior of covariance matrices $\alpha \boldsymbol{\Sigma}_{\mathbf{d}}$ characterizing a bi-dimensional fBm of Hurst exponent $H=1$, with uncorrelated bi-variate components, truncated on the image grid $\Omega_m$. 
 The stack of images $\xx^\star_{t_0}$ is computed using a ground truth displacement field $\mathbf{d}^\star$ and   ground truth reference images $\xx^\star_{t_1}$ according to the warping model \eqref{eq:warpingModel}.  We set the ground truth displacement as a realization of the prior model, while
 the stack of images $\xx^\star_{t_1}$ is chosen to be real-world images taken from the  ECMWF numerical simulation.
 We use a couple of grids $(\Omega^{t_0}_{obs},\Omega^{t_1}_{obs})$ taken from real-world IASI partial observations to generate observations  $\yy$ in a noise-free setting: $\yy_t^{obs}(s)=\xx^\star_{t}(s)$ for $s\in \Omega^{t}_{obs},\, t=t_0,t_1$.   
 
 For experiment \#2,   the target posterior distribution is the same (up to different parameter $\alpha,\gamma$) as for experiment \#1 .  However the observations $\yy$  are different: they are  defined as the output of the simulation of the  ECMWF numerical model, in a noise-free setting  on the entire grid $\Omega^{t_0}_{obs}=\Omega^{t_1}_{obs}=\Omega_m$. 
 
  For experiment \#3,    the target posterior distribution remains unchanged (up to different parameter $\alpha,\gamma$), while  observations $\yy$ differ.  The incomplete observations $\yy_t^{obs}(s)$
  for $s\in \Omega^{t}_{obs},\, t=t_0,t_1$  are provided in this last challenging case by  real-world IASI  measurements. A \textit{proxy} to the ground truth is given by the operational numerical weather model simulated at the corresponding   time and locations.\medskip

 In all these experiments, the  time discretization step $\delta t$  in the random walks, the MALA or  the HMC algorithm are manually tuned  together with the number of leap-frog iterations $L$ used for HMC, with the objective  to reach an acceptance rate around $0.9$.

\subsection{Numerical results}

\begin{figure}[!h]
\begin{center}\vspace{-0.cm}
\begin{tabular}{cc|cc}
\hspace{-0.5cm}\includegraphics[width=0.23\textwidth]{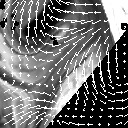}&\hspace{-0.25cm}\includegraphics[width=0.23\textwidth]{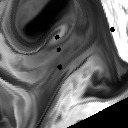}&\hspace{-0.05cm}\includegraphics[width=0.23\textwidth]{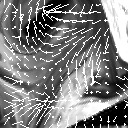}&\hspace{-0.25cm}\includegraphics[width=0.23\textwidth]{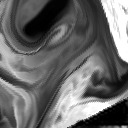}\\
\multicolumn{2}{c}{\footnotesize{True displacement $\mathbf{d}^\star$ and images $\mathbf{y}^{1,obs}_{t_0},\mathbf{y}^{1,obs}_{t_1}$}}&\multicolumn{2}{c}{\footnotesize{MAP displacement $ \mathbf{\hat d}$ and images  $ \mathbf{\hat x}^1_{t_0}, \mathbf{\hat x}^1_{t_1}$.}}
\end{tabular}
	\caption{{\footnotesize  {  \textbf{Experiment \#2.} Left: true AMVs superimposed on the pair of incomplete image observations (black pixels correspond to missing data). Right: comparison to the MAP estimate.  \label{fig:1}}}}
		\end{center}\vspace{-0.2cm}
\end{figure}

\subsubsection{ Experiment\#1:  synthetic turbulence}
The first layer   of the observed synthetic partial image stack with the superimposed ground truth displacement field are displayed in Figure~\ref{fig:1}.   They are to be compared to the MAP estimate obtained by a deterministic gradient-based method.
 The table below presents the performances  of the different methods (with $N\times L$=$1e3$, $L$=$10$, $H$=$0.5$) in terms of the endpoint error criteria \eqref{eq:MSE} -- \eqref{eq:weightL0_}  presented in the Section~\ref{sec:evalCriteria}. 
\begin{center}
\begin{tabular}{c||c|c|c||c|c||c}
&\multicolumn{6}{c}{Endpoint Error: $EPE(\Omega,w,\boldsymbol{\theta^\star}-  \boldsymbol{\hat \theta})$ }\\
\hline
\hspace{-1.cm} $(\Omega,w)$= & $(\Omega_m,\mathbf{1})$ & $(\Omega_m,w^1)$ &$(\Omega_m,w^2)$ & $(\Omega_{obs},\mathbf{1})$  &  $(\Omega_m,w^0)$   &  $(\Omega_{obs},w^0)$   \\
\hline
\hline
\hspace{-1.cm}  Laplace   &0.655976 &0.620422&0.597836 & 0.475874&0.519058&0.329363\\
\hspace{-1.cm}    Stand. RW & 0.656146 & 0.685375& 0.714258 & 0.476116&0.653051&0.470416\\
\hspace{-1.cm}    Precond. RW& 0.656935 & 0.702102& 0.753994 & 0.479103&0.68768&0.518113\\
\hspace{-1.cm}    MALA ($\zeta$=1)& 0.828738 & 0.813179& 0.795965 & 0.66471&0.758065&0.583866\\
\hspace{-1.cm}    HMC ($\zeta$=1) & 1.27958 & 1.20653& 1.13141 & 1.04521&1.06443&0.797037\\
\hspace{-1.cm}  MALA  ($\zeta$=1e-6) &0.597302 &0.571421 & 0.597154&0.41554&0.448114&0.230732\\
\hspace{-1.cm} HMC  ($\zeta$=1e-6)  &\textbf{0.452115} & \textbf{0.391389} & \textbf{0.295442}&\textbf{0.291273}& \textbf{0.295208}&\textbf{0.172352}\\
 \end{tabular}\vspace{0.2cm}
\end{center}

A first glance leads us to notice that the Laplace method provides  bad approximations (with a computation time more than 2 times longer),  suggesting a poor numerical approximation of the log posterior Hessian and of its inverse (detailed in Appendix~\ref{rem:CondIndLaplace}). Random walks  achieve inaccurate approximations as well. 
We also remark that using the standard MALA  algorithm without chilling  can substantially  degrade the various EPE criteria.  This undesirable behavior is even more prominent for the standard HMC algorithm.  In contrast, chilling brings a significant  gain in terms of the different EPE criteria, and is particularly a powerful strategy when combined with the HMC algorithm: chilled HMC reaches a gain of  more than $30\%$ in terms of the standard EPE, and this gain is increased to $50\%$ considering weighting functions  $ \eqref{eq:weightL1_} $ for $p=1,2$.  Sparse EPE and masked EPE are comparable, with a gain of nearly  $40\%$ with respect to Laplace.  The sparse masked EPE (using weights  $ \eqref{eq:weightL0_} $) yields to a large error decrease: about $50\%$ lower than its Laplace counterpart. Let us finally remark that there exists a large gap  in accuracy between the standard EPE on the entire domain $\Omega_m$ computed for the MAP estimate, and the sparse masked EPE revealing the average error on a selected subdomain of $\Omega_{obs}$ , which is more than $70\%$ more accurate. 

\begin{figure}[!h]
\begin{center}
\begin{tabular}{cc}
\vspace{-0.35cm}\includegraphics[width=0.5\textwidth]{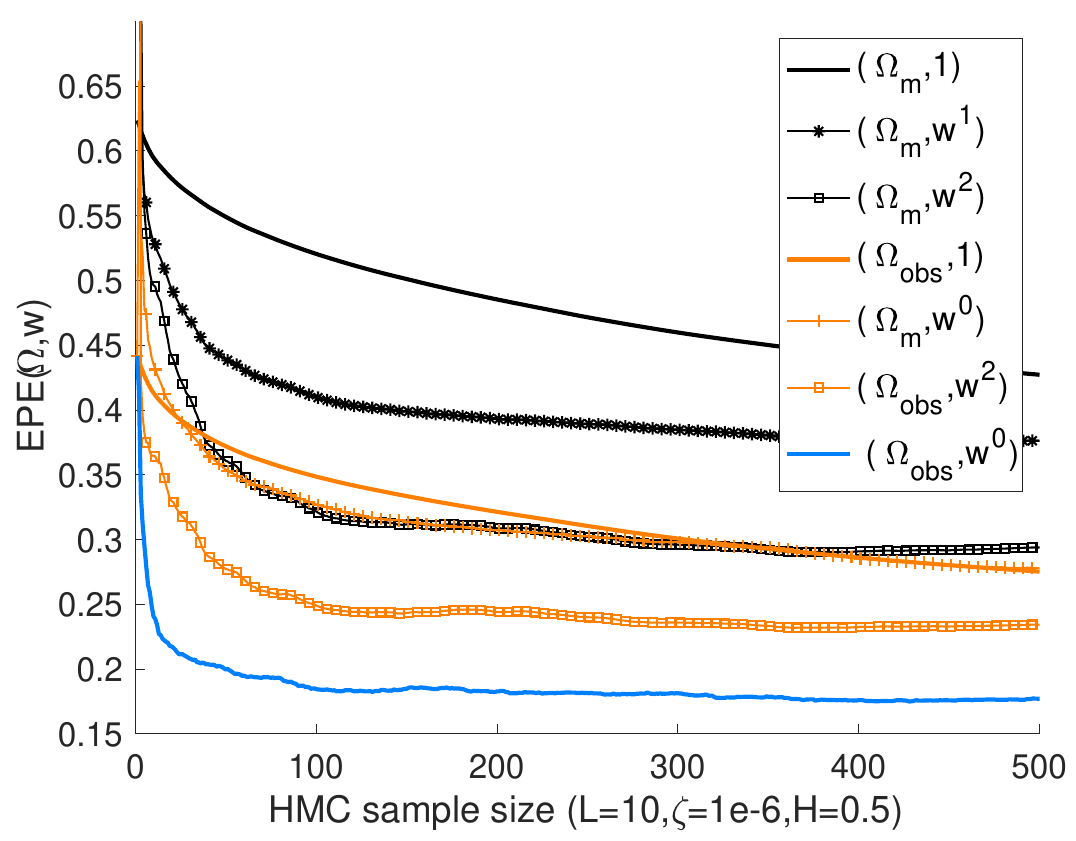}
\end{tabular}
	\caption{{\footnotesize  {  \textbf{Experiment \#1.} Comparison of the various  endpoint error criteria \eqref{eq:MSE} -- \eqref{eq:weightL0_} with respect to the sample size ($N\times L$) for a chilled HMC simulation. \label{fig:2}}}}
		\end{center}
\end{figure}

\begin{figure}[!h]
\begin{center}
\begin{tabular}{cc}
\vspace{-0.35cm}\hspace{-1.cm}\includegraphics[width=0.5\textwidth]{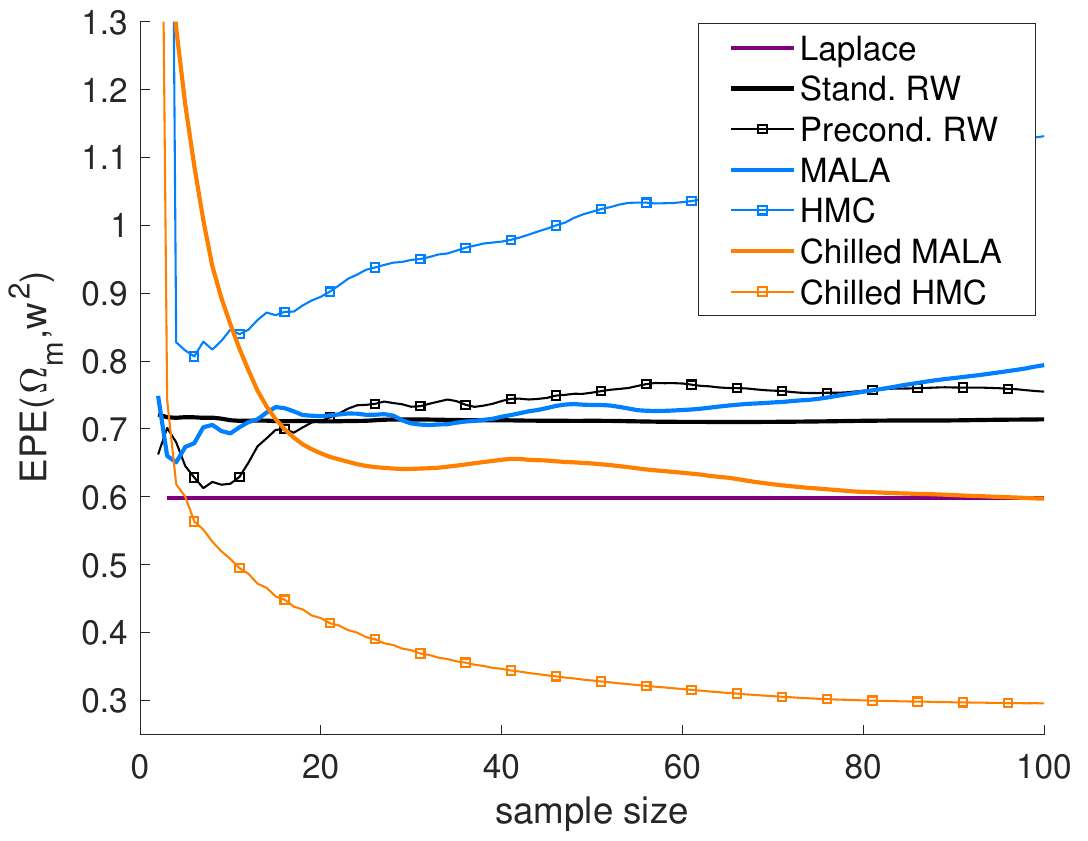}&\hspace{-0.45cm}\includegraphics[width=0.5\textwidth]{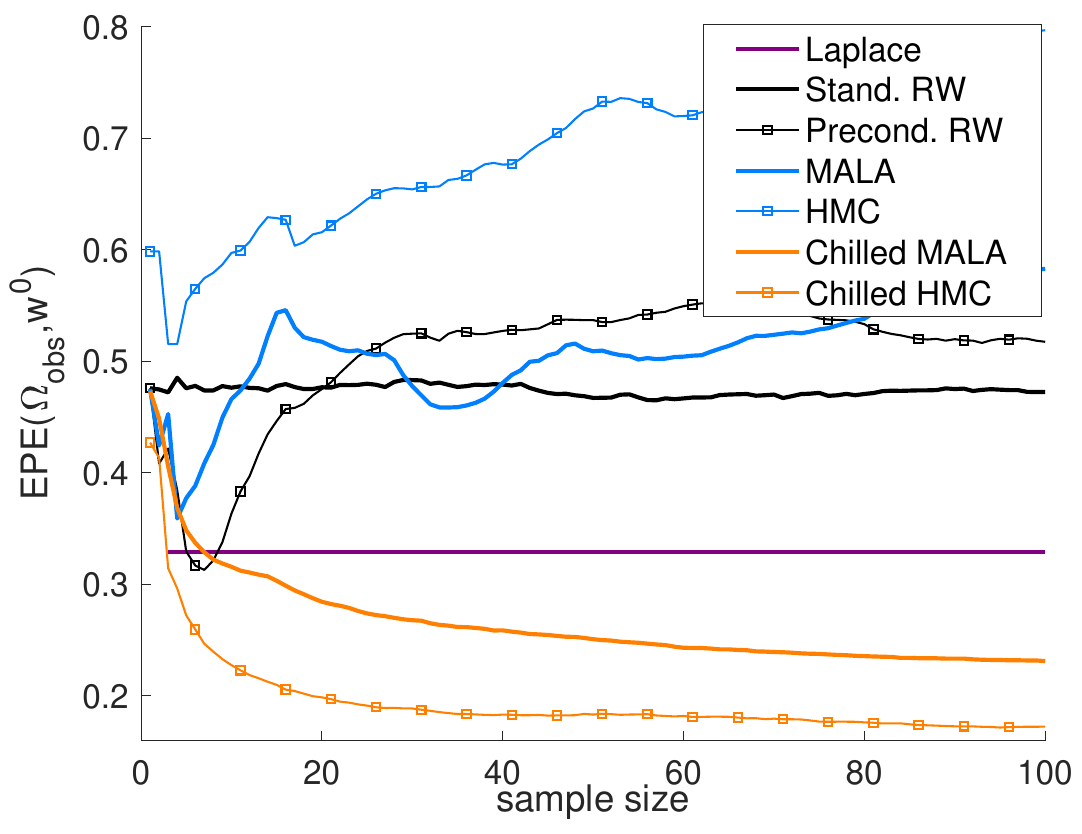}
\end{tabular}
	\caption{{\footnotesize  {  \textbf{Experiment \#1.} Comparison of the different methods in terms of the endpoint error criteria \eqref{eq:MSE} with \eqref{eq:weightL1_} for $p$=2 (left)  or with \eqref{eq:weightL0_} (right) with respect to the sample size $N\times L$=1e3  ($H$=0.5,$\zeta$=1e-6).  \label{fig:3}}}}
		\end{center}
\end{figure}

\begin{figure}[!h]
\begin{center}
\begin{tabular}{cc}
\vspace{-0.5cm}\hspace{-1.cm}\includegraphics[width=0.5\textwidth]{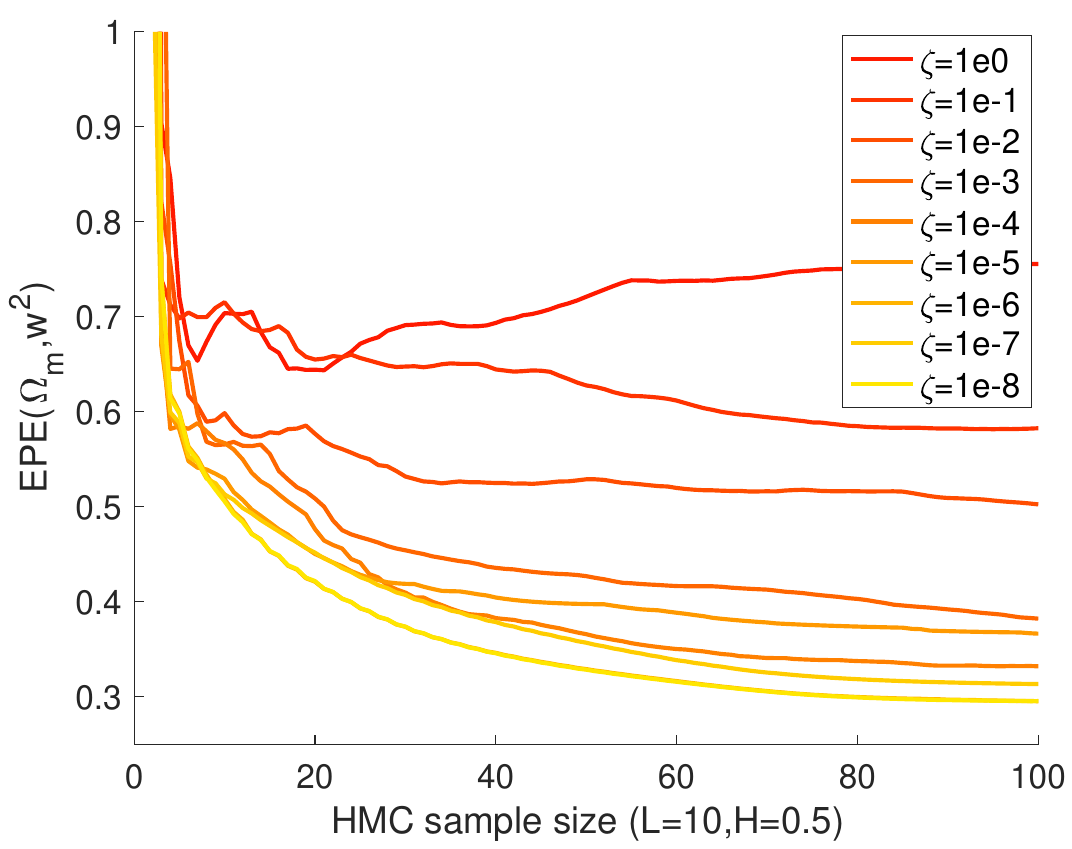}&\hspace{-0.45cm}\includegraphics[width=0.5\textwidth]{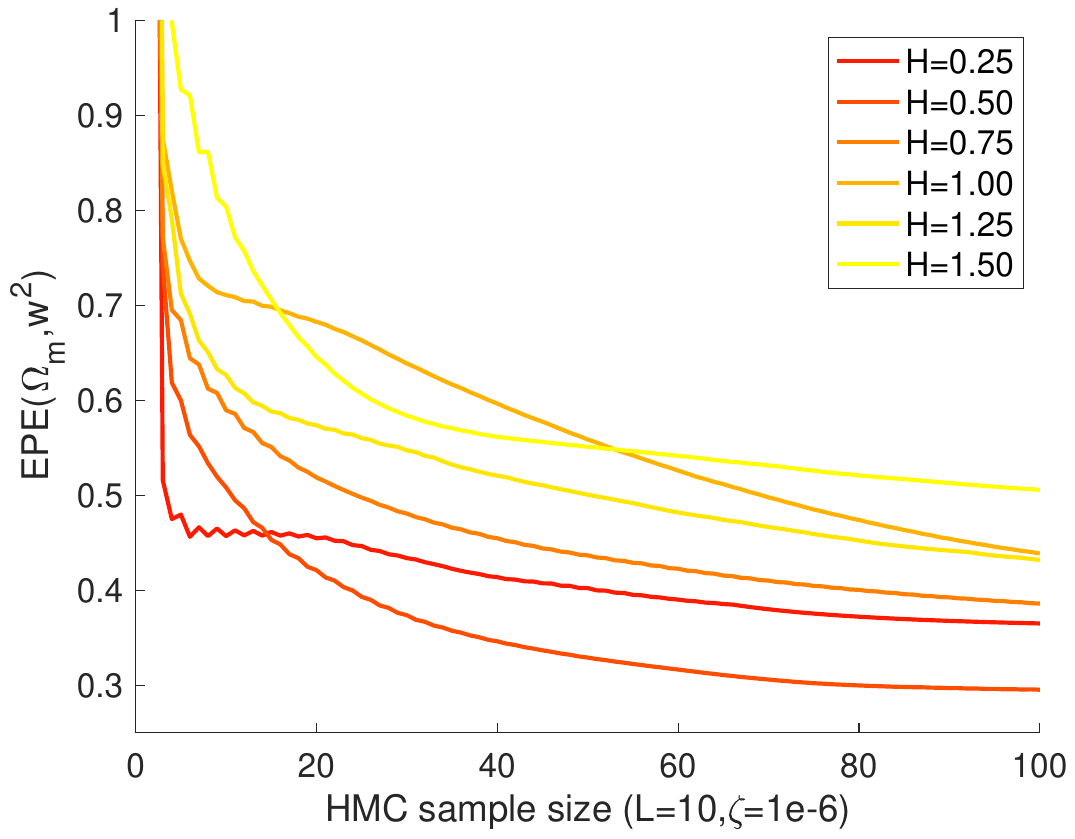}
\end{tabular}
	\caption{{\footnotesize  {  \textbf{Experiment \#1.} Influence of the temperature $\zeta$ (left) and of the preconditioning parameter $H$ (right) on the evolution of  the endpoint error criteria \eqref{eq:MSE} with \eqref{eq:weightL1_} for $p$=2. \label{fig:4}}}}
		\end{center}\vspace{-0.cm}
\end{figure}

Figure~\ref{fig:2} displays the  evolutions of the EPE criteria with respect to the sample size of the chilled HMC simulation. As expected the $p$-weighted criteria converge  to lower values than the standard EPE, and moreover they converge faster.  Besides, we observe  that the sparse and masked EPE achieve  comparable performances.
 Figure~\ref{fig:3} shows the evolution of 2-weighted EPE and the sparse EPE criteria for the different methods. We observe that the chilled algorithms are the only ones which yield stable and low EPE criteria. 
Figure~\ref{fig:4} shows the influence of the chilling and preconditioning parameter with respect to the  2-weighted EPE criterion. We observe on the one hand that a Hurst exponent equal to half of the one of the prior yields the best results, and on the other hand that the lower the temperature, the faster and more stable the convergence of this criterion. 

\begin{figure}[!t]
\begin{center}
\begin{tabular}{cc||cc}
\hline 
\multicolumn{2}{c}{\footnotesize{Laplace}}&\multicolumn{2}{c}{\footnotesize{Chilled HMC}}\\
\hline \\
\hspace{-0.5cm}\includegraphics[width=0.23\textwidth]{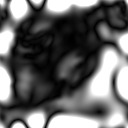}&\hspace{-0.25cm}\includegraphics[width=0.23\textwidth]{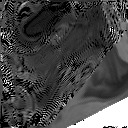}&\hspace{-0.05cm}\includegraphics[width=0.23\textwidth]{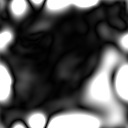}&\hspace{-0.25cm}\includegraphics[width=0.23\textwidth]{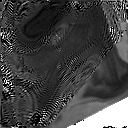}\\
{\footnotesize{$\|\mathbf{d}^\star(s)- \mathbf{\hat d}(s)\|_2$}}&{\footnotesize{$|\mathbf{x}_{t_1}^{1,\star}(s)- \mathbf{\hat x}^1_{t_1}(s)|$}}&{\footnotesize{$\|\mathbf{d}^\star(s)- \mathbf{\hat d}(s)\|_2$}}&{\footnotesize{$|\mathbf{x}_{t_1}^{1,\star}(s)- \mathbf{\hat x}^1_{t_1}(s)|$}}\\
\hspace{-0.5cm}\includegraphics[width=0.23\textwidth]{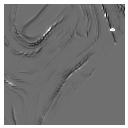}&\hspace{-0.25cm}\includegraphics[width=0.23\textwidth]{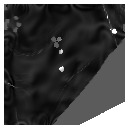}&\hspace{-0.05cm}\includegraphics[width=0.23\textwidth]{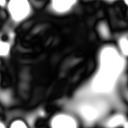}&\hspace{-0.25cm}\includegraphics[width=0.23\textwidth]{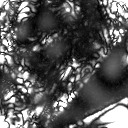}\\
\multicolumn{2}{c}{\footnotesize{${\mathcal{E}}_{\hat \mu,\boldsymbol{\hat \theta}}(\psi)$, left:  $\psi(\theta)$=$\mathbf{d}(s)$, right: $\psi(\theta)$=$\mathbf{x}_{t_1}^1(s)$}}  &\multicolumn{2}{c}{\footnotesize{${\mathcal{E}}_{\hat \mu,\boldsymbol{\hat \theta}}(\psi)$, left:  $\psi(\theta)$=$\mathbf{d}(s)$, right: $\psi(\theta)$=$\mathbf{x}_{t_1}^1(s)$}} \vspace{-0.2cm}
\end{tabular}
	\caption{{\footnotesize  {  \textbf{Experiment \#1.} True versus expected errors obtained by the Laplace method or with chilled HMC (N=1e2, L=10, H=0.5, $\zeta$=1e-6). The gray level values range in $[0,l_{\max}]$, $l_{\max}$ being equal to  the empirical mean plus standard deviation over $\Omega_m$ of the true or expected errors (the values of $\|\mathbf{d}^\star(s)- \mathbf{\hat d}(s)\|_2$  are  thresholded to $l_{\max}\approx 1.3$). \label{fig:5}}}}
		\end{center}\vspace{-0.5cm}
\end{figure}

\begin{figure}[!t]
\begin{center}
\begin{tabular}{cccc}
\hline 
\footnotesize{Laplace}&\footnotesize{Standard RW}&\footnotesize{Precond. RW}&\footnotesize{MALA}\\
\hline 
\hspace{-0.5cm}\rotatebox{90}{\parbox{2mm}{\multirow{1}{*}{{\footnotesize{$\quad\quad\|\mathbf{d}^\star(s)- \mathbf{\hat d}(s)\|_2$}}}}}\hspace{0.25cm}\includegraphics[width=0.23\textwidth]{./Images/Experiment2/Laplace/UV_trueL2ErrorMAP.jpeg} &\hspace{-0.25cm}\includegraphics[width=0.23\textwidth]{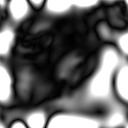}&\hspace{-0.25cm}\includegraphics[width=0.23\textwidth]{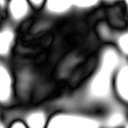}&\hspace{-0.05cm}\includegraphics[width=0.23\textwidth]{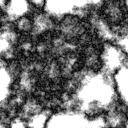}\\
\hspace{-0.5cm}\rotatebox{90}{\parbox{2mm}{\multirow{1}{*}{\footnotesize{$\quad\quad\mathbf{\hat d}$ and ${\mathcal{E}}_{\hat \mu,\boldsymbol{\hat \theta}}(\mathbf{ d}(s))$}}}}\hspace{0.25cm}\includegraphics[width=0.23\textwidth]{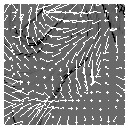}& \hspace{-0.25cm}\includegraphics[width=0.23\textwidth]{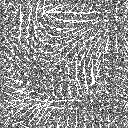}&\hspace{-0.25cm}\includegraphics[width=0.23\textwidth]{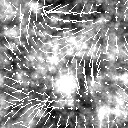}&\hspace{-0.05cm}\includegraphics[width=0.23\textwidth]{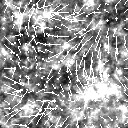}\\
\hline 
\footnotesize{HMC}&\footnotesize{Chilled MALA}&\footnotesize{Chilled HMC}&\footnotesize{Missing obs. and truth}\\
\hline 
\hspace{-0.5cm}\rotatebox{90}{\parbox{2mm}{\multirow{1}{*}{{\footnotesize{$\quad\quad\|\mathbf{d}^\star(s)- \mathbf{\hat d}(s)\|_2$}}}}}\hspace{0.25cm}\includegraphics[width=0.23\textwidth]{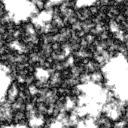}&\hspace{-0.25cm} \includegraphics[width=0.23\textwidth]{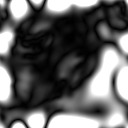}&\hspace{-0.25cm}\includegraphics[width=0.23\textwidth]{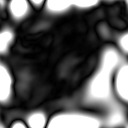}&\rotatebox{90}{\parbox{2mm}{\multirow{1}{*}{{\footnotesize{ $\quad\quad\quad\quad\overline{\Omega}_{obs}$}}}}}\hspace{0.cm}\hspace{-0.05cm}\includegraphics[width=0.23\textwidth]{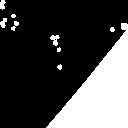}\\
\hspace{-0.5cm}\rotatebox{90}{\parbox{2mm}{\multirow{1}{*}{\footnotesize{$\quad\quad\mathbf{\hat d}$ and ${\mathcal{E}}_{\hat \mu,\boldsymbol{\hat \theta}}(\mathbf{ d}(s))$}}}}\hspace{0.25cm}\includegraphics[width=0.23\textwidth]{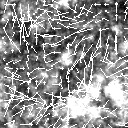}& \hspace{-0.25cm}\includegraphics[width=0.23\textwidth]{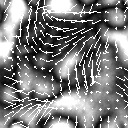}&\hspace{-0.25cm}\includegraphics[width=0.23\textwidth]{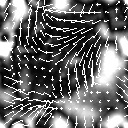}&\rotatebox{90}{\parbox{2mm}{\multirow{1}{*}{{\footnotesize{$\quad\quad\quad\quad\quad\mathbf{d}^\star$}}}}}\hspace{0.cm}\hspace{-0.05cm}\includegraphics[width=0.23\textwidth]{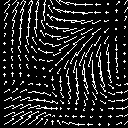}\vspace{-0.3cm}
\end{tabular}
	\caption{{\footnotesize  {  \textbf{Experiment \#1.}  True versus expected errors  obtained with the different methods.  AMV estimates are superimposed on the expected errors. 
	The gray level values range in $[0,l_{\max}]$, $l_{\max}$ being equal to  the empirical mean plus standard deviation over $\Omega_m$ of the true or expected errors (the values of $\|\mathbf{d}^\star(s)- \mathbf{\hat d}(s)\|_2$  are  thresholded to $l_{\max}\approx 1.5$ and $2.3$ for standard MALA and HMC and $\approx 1.3$ for the other algorithms). \label{fig:6}}}}
		\end{center}\vspace{-0.2cm}
\end{figure}

Let us draw from the table and these plots a first set of conclusion. First, chilling accelerates the characterization of the expected error, which converges within a limited simulation budget.  In particular, the lower the temperature, the more accurate the estimates. Second, the combination of HMC and chilling is performant: 1) PM estimate provided by the chilled version of HMC is  accurate, and in particular it is substantially more accurate than the MAP estimate; 2) the convergence of the expected error estimate is substantially accelerated by this combination, as attested by the $p$-weighted  EPE criteria; 3) the Hurst exponent used for preconditioning is not necessarily the one corresponding to the prior.  Third,  sparse and masked EPE are  comparable, suggesting that  much of the large expected errors are concentrated on non-observed areas.  Last, the subset of AMVs selected by the weighting function  $ \eqref{eq:weightL0_} $  is expected to have very good accuracy, and is indeed reliable.

Figure~\ref{fig:5} compares the spatial distribution of the true and expected error norms for the displacement variable and the image variable (we only display the error related to the first layer of the stack of images, \ie, for  $\ell$=$1$).  Although the expected image error matches the true one,  the expected displacement errors obtained by  the Laplace method appear to be closely related to the spatial derivatives of the image observations. These structures are spatially very localized and not relevant of the general structure of the errors.  This suggests that the numerical approximations performed to construct and invert the Hessian presented in Appendix~\ref{rem:CondIndLaplace} are insufficient to accurately account for the spatial dependencies of the displacement variable. Chilled HMC  provides an enhancement  through more accurate estimation of errors in a large range of scales. We note that the large-scale structures of errors are mainly recovered in regions related to missing observations or borders.  Figure~\ref{fig:6} compares the spatial distribution of the true and expected error norms for the displacement variable, on which is superimposed the displacement estimated mean.  We verify visually that the standard random walk exhibits a very slow convergence:  it yields an expected error close to white noise, while the estimated mean remains nearly equal to the Laplace estimate. Substituting  the diagonal covariance for  the fBm covariance in the random walk leads to a  structured but irrelevant spatial distribution of the error. Conversely, the standard MALA and HMC error estimates contain most of the major spatial structures of the true   error,  but these are embedded in significant noise. Moreover,  the displacement mean estimates are significantly corrupted by noise. The chilled version of the MALA and HMC  algorithms show a striking refinement on the characterization of the spatial distribution of the error of the PM estimates. We can clearly identify the mid-scale structures of the true error.

\begin{figure}[!t]
\vspace{-0.cm}\begin{center}
\begin{tabular}{cc||cc}
\multicolumn{2}{c}{\footnotesize  {  ECMWF  model  }}&\multicolumn{2}{c}{\footnotesize  {  IASI  observations (synchronized)}}\\
\hline \\
\hspace{-0.5cm}\rotatebox{90}{\parbox{2mm}{\multirow{1}{*}{{\footnotesize{$\quad\quad \mathbf{y}^{1,obs}$ and $\mathbf{d}^\star$}}}}}\hspace{0.25cm}\includegraphics[width=0.23\textwidth]{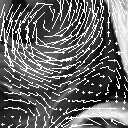}&\hspace{-0.25cm}\includegraphics[width=0.23\textwidth]{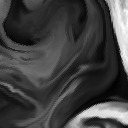}&\hspace{-0.cm}\includegraphics[width=0.23\textwidth]{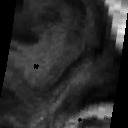}&\hspace{-0.25cm}\includegraphics[width=0.23\textwidth]{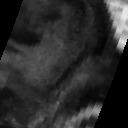}\\
\hspace{-0.5cm}\rotatebox{90}{\parbox{2mm}{\multirow{1}{*}{{\footnotesize{$\quad\quad \quad\quad \mathbf{y}^{2,obs}$}}}}}\hspace{0.25cm}\includegraphics[width=0.23\textwidth]{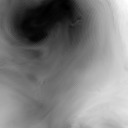}&\hspace{-0.25cm}\includegraphics[width=0.23\textwidth]{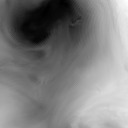}&\hspace{-0.cm}\includegraphics[width=0.23\textwidth]{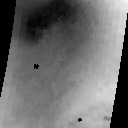}&\hspace{-0.25cm}\includegraphics[width=0.23\textwidth]{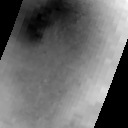}\\
\hspace{-0.5cm}\rotatebox{90}{\parbox{2mm}{\multirow{1}{*}{{\footnotesize{$\quad\quad \quad\quad \mathbf{y}^{3,obs}$}}}}}\hspace{0.25cm}\includegraphics[width=0.23\textwidth]{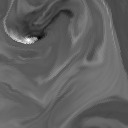}&\hspace{-0.25cm}\includegraphics[width=0.23\textwidth]{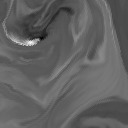}&\hspace{-0.cm}\includegraphics[width=0.23\textwidth]{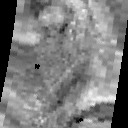}&\hspace{-0.25cm}\includegraphics[width=0.23\textwidth]{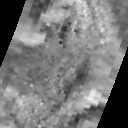}\\
\footnotesize{$t_0$}&\footnotesize{$t_1$}&\footnotesize{$t_0$}&\footnotesize{$t_1$}\vspace{-0.2cm}
\end{tabular}
	\caption{{\footnotesize  {  \textbf{Experiment \#2 \& \#3.} AMVs from the ECMWF  simulation and  the pairs of image observations (black pixels correspond to missing data): humidity (above), temperature (middle), ozone concentration (below).  \label{fig:7}}}}\vspace{-0.cm}
		\end{center}\vspace{-0.5cm}
\end{figure}

Visual inspection of these error maps corroborates our previous findings, another qualitative one being that the standard and chilled MCMC simulations appear to converge to similar estimates, although the chilled counterparts are much faster. We observed in our experiments that convergence is accelerated by about $\times $25 using the proposed chilled approach, a proxy for the speed of convergence being  equated to the number of samples needed to go below a given weighted MSE value.

\subsubsection{Experiments \#2 \& \#3: real-world observations}

We now turn to a realistic meteorological context. In experiments \#2, the image observations (available for all points of the  grid  $\Omega_m$) of humidity, temperature and ozone concentration are  generated by the ECMWF numerical simulation. They are displayed on the left of  Figure~\ref{fig:7}, together with the related ground truth horizontal wind field.    In experiments \#3, the image observations are synchronized IASI observations provided by the satellites. They are displayed on the right  of the same  figure.

The table below presents the performances  of the different methods ($N\times L$=1e3, $L$=$10$, $H$=0.5) in terms of the  error criteria \eqref{eq:MSE} -- \eqref{eq:weightL0_} for experiment \#2. 

\begin{center}
\begin{tabular}{c||c|c|c||c}
&\multicolumn{4}{c}{Endpoint Error: $EPE(\Omega,w,\boldsymbol{\theta^\star}-  \boldsymbol{\hat \theta})$ }\\
\hline
\hspace{-1.cm}   $(\Omega,w)$= & $(\Omega_m,\mathbf{1})$ & $(\Omega_m,w^1)$ &$(\Omega_m,w^2)$  &  $(\Omega_{m},w^0)$   \\
\hline
\hline
\hspace{-1.cm}  Laplace   &0.721005 &0.77865&\textbf{0.742759} &0.736874 \\
\hspace{-1.cm}    Stand. RW & 0.720978 & 0.758557& 0.796114 & 0.728775\\
\hspace{-1.cm}    Precond. RW& 0.718858 & 0.732256& {0.744423} & 0.668264\\
\hspace{-1.cm}    MALA ($\zeta$=1)& 0.71656 & 0.739782& 0.762355 & 0.711341\\
\hspace{-1.cm}    HMC ($\zeta$=1) & 0.713046 & 0.731385& 0.752246 & 0.676624\\
\hspace{-1.cm}  MALA  ($\zeta$=1e-2) &0.712728 &0.757558 & 0.78937&0.649664\\
\hspace{-1.cm} HMC  ($\zeta$=1e-2)  &\textbf{0.688351} & \textbf{0.725055} & {0.774966}& \textbf{0.617561}
 \end{tabular}
	\end{center}\vspace{0.2cm}

We find the same trend as for the synthetic  turbulence data in terms of standard and sparse EPE criteria: the chilled HMC outperforms the other methods with a reduction in standard and sparse EPE of about 5\% and 15\%, respectively, compared to the Laplace approximation. However, unlike the sparse EPE criterion, the $p$-weighted EPE criteria remain high using a chilled HMC, and in particular the use of a constant weighting function yields lower criteria. We deduce that the accurate estimation of the expected error over the entire domain seems to be difficult to achieve in real-world scenarios. Indeed,  unlike the case where the ground truth has been generated by the prior (the case of experiment \#1),  the prior is in this case a simplification of the real-world atmospheric dynamics, and provides only up to some extent  relevant information on the probability structure of the displacement field. Nevertheless, the results suggest that regions characterized by low expected errors can still be well discriminated.

We finally assess the performances  of the different methods in the context of  experiment \#3, \ie, for  the  case of IASI incomplete and noisy observations. The table below details these performances and compare them to a state-of-the-art optic-flow algorithm for AMV estimation~\cite{Heas14,borde2019winds}. 
\begin{center}
\begin{tabular}{c||c|c|c||c|c||c}
&\multicolumn{6}{c}{Endpoint Error: $EPE(\Omega,w,\boldsymbol{\theta^\star}-  \boldsymbol{\hat \theta})$ }\\
\hline
\hspace{-1.cm}   $(\Omega,w)$= & $(\Omega_m,\mathbf{1})$ & $(\Omega_m,w^1)$ &$(\Omega_m,w^2)$ & $(\Omega_{obs},\mathbf{1})$  &  $(\Omega_m,w^0)$   &  $(\Omega_{obs},w^0)$   \\
\hline
\hline
\hspace{-1.cm}  Algorithm~\cite{borde2019winds}& 1.62478&-&-&1.59516&-&-\\
\hspace{-1.cm}  Laplace   &{1.29478} &\textbf{1.30341}&\textbf{1.29047} & {1.24146}&\textbf{1.2394}&1.2638\\
\hspace{-1.cm}    Stand. RW & 1.2948 & 1.35004& {1.40486} & 1.24148&1.29575&1.24388\\
\hspace{-1.cm}    Precond. RW& 1.29500& 1.41510 & 1.49525 & \textbf{1.24118}&1.32099&1.26990\\
\hspace{-1.cm}    MALA ($\zeta$=1)& \textbf{1.29474} & 1.37928& 1.45494 & 1.24182&1.32444&1.3306\\
\hspace{-1.cm}    HMC ($\zeta$=1) & 1.29955 & 1.36311& 1.43135 & 1.24546&1.30417&1.30207\\
\hspace{-1.cm}  MALA  ($\zeta$=1e-4) &1.29799 &1.50883 & 1.76603&1.24278&1.26168&1.26478\\
\hspace{-1.cm} HMC  ($\zeta$=1e-4)  &{1.30440} & {1.42415} & {1.53516}&{1.2479}& {1.27541}&\textbf{1.16574}
 \end{tabular}\vspace{0.2cm}
\end{center}

Before discussing these results, let us underline  major  difficulties making the context of experiment \#3 very challenging: 1) the presence of strong noise whose distribution is physically difficult to characterize; 2) the presence of regions with missing observations;  3) the real-world atmospheric dynamics that can strongly deviate from our simplified posterior model. Moreover, the real-world atmospheric dynamics which have generated the IASI observations may also significantly deviate from the ECMWF numerical simulation.  Therefore, the evaluation is critical because we cannot really rely on a ground truth horizontal wind field, but only to some extent on the ECMWF numerical simulation. Nevertheless, in the absence of any other ground truth, we use  the numerical weather simulation as a reference. 

Based on the reliability of the state-of-the-art algorithm~\cite{borde2019winds}, the table above shows that the different methods provide relatively accurate estimates of the displacement field in terms of standard EPE. Indeed, we observe a gain of around $20\%$ provided by one or other of the methods evaluated compared to this algorithm. But, in general, we observe that the expected errors calculated by the Laplace method or almost all the MCMC methods do not contribute much to reducing the standard EPE. Indeed, when inspecting the various EPE criteria, we notice a few small improvements that look more like the effect of chance, without any of the methods evaluated being clearly superior to the other.
Nevertheless, one exception stands out: the $p$-weighted EPE criteria for $p=0$ and restricted to the observed domain provided by the chilled HMC simulation.   We observe in this case a clear gain of over $10\%$ on the EPE criterion, compared with standard EPE. These results show that the regions associated with low values of the expected error estimated by the chilled HMC are on average consistent with the ground-truth error. Once again, however, it is difficult to draw quantitative conclusions, as the interpretation of these results must be tempered by the fact that we are referring to the ECMWF simulation, as the ground truth is unknown.

\begin{figure}[!t]
\begin{center}
\vspace{-0.5cm}\begin{tabular}{cc||cc}
\hline 
\multicolumn{2}{c} {\footnotesize  Experiment\#2 } & \multicolumn{2}{c} {\footnotesize   Experiment\#3 } \\
\hline \\
\hspace{-0.cm}\includegraphics[width=0.23\textwidth]{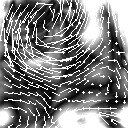}&\hspace{-0.25cm}\includegraphics[width=0.23\textwidth]{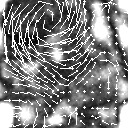} &\hspace{-0.cm}\includegraphics[width=0.23\textwidth]{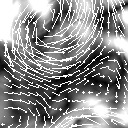}&\hspace{-0.25cm}\includegraphics[width=0.23\textwidth]{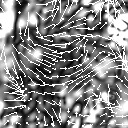}\\
{\footnotesize  $ \mathbf{d}^\star$ and $\|\mathbf{d}^\star(s)-\mathbf{\hat d}(s)\|_2$  } & {\footnotesize  $ \mathbf{\hat d}$ and  ${\mathcal{E}}_{\hat \mu,\boldsymbol{\hat \theta}}(\mathbf{ d}(s))$  } & {\footnotesize  $ \mathbf{d}^\star$ and $\|\mathbf{d}^\star(s)-\mathbf{\hat d}(s)\|_2$  } & {\footnotesize  $ \mathbf{\hat d}$ and  ${\mathcal{E}}_{\hat \mu,\boldsymbol{\hat \theta}}(\mathbf{ d}(s))$  }  \vspace{-0.2cm}
\end{tabular}
	\caption{{\footnotesize  {   \textbf{Experiment\#2 \& \#3.} {True versus expected error obtained by chilled HMC for experiment\#2 (left) and \#3 (right).  AMV ground truth (resp. estimate) are superimposed on the true error (resp. expected error). The gray level values range in $[0,l_{\max}]$, $l_{\max}$ being equal to  the empirical mean plus standard deviation over $\Omega_m$ of the true or expected errors (the values of the true error  are  thresholded to $l_{\max}\approx 1.0$  (resp. $\approx 1.9$)  for experiment \#2  (resp. experiment \#3)).}  \label{fig:8}}}}\vspace{-0.2cm}
		\end{center}\vspace{-0.2cm}
\end{figure}

The spatial distribution of the true errors  can be compared to the expected error estimates obtained by chilled HMC in Figure~\ref{fig:8}  for experiments \#2 and \#3 (the ECMWF simulation is assumed to be the ground truth in both experiments). The figure also compares the true and the estimated displacement field overlaid on these error maps. We observe that a majority of the error structures at different scale ranges have been fairly characterized in experiments \#2.    One may point out a contradiction with the poor estimation of the expected error,  revealed by  the high  $p$-weighted EPE criteria in the related table.  The latter criteria reveal indeed the low quality of the uncertainty estimates but the criteria remain nevertheless very sensitive to strong local deviations of the expected error estimates, although these deviations may be restricted to certain regions. Regarding experiment \#3, we note that the estimates are globally consistent with the ECMWF reference, although there are some structural differences between the actual displacement field and that estimated by our chilled HMC algorithm.  The differences between the spatial distribution of true and expected errors are not easily interpreted. Although close inspection reveals that the displacement fields and errors show some local similarities, we are led to believe that the ECMWF simulation may deviate locally from the ground truth data, and is therefore a tricky benchmark for evaluation, preventing any strong conclusions from being drawn. 

\section{Conclusions}
 The starting point of this work is to address the crucial need to accurately characterize the uncertainty in deterministic AMV estimates prior to feeding numerical weather prediction models. The paper provides an innovative framework for refining the estimation of the  AMVs, while estimating jointly their expected  errors. We show that an effective method is to sample a low-temperature approximation of the posterior distribution (related to Bayesian AMV model), using an HMC algorithm preconditioned by the covariance of the prior itself -- an isotropic fBm -- but with possibly differently fitted hyper-parameters. 

 From a more general and theoretical point of view, we prove that lowering the temperature actually amounts to approximating a local Gaussian approximation of the posterior around a point estimate and in the zero temperature limit, the chilled distribution converges in distribution to the Laplace point approximation, which is in general out of reach  for high-dimensional non-Gaussian posterior.  
 
Several numerical  benchmarks (with associated ground truth data)  are designed to evaluate the algorithm performances: data sets range from a toy-model simulation to a real ECMWF weather simulation and to IASI meteorological satellite images.  In order to quantitatively judge the relevance of the AMV estimates, we  introduce a family of error criteria. These criteria minimize some Bayesian risk under various specific structural constraints on the weights.   These  criteria show that the chilled MCMC simulation significantly increases the accuracy of the point estimates of AMVs and their expected errors. In particular, we observe an improvement over the standard  estimates or our deterministic numerical approximation of the Laplace estimate. Moreover, the significant decrease of these criteria in a short time shows that chilling significantly accelerates the convergence speed of the algorithm. 
 
Beyond the case of Gaussian posteriors, this work raises the open question of determining sufficient conditions ensuring the quality of the proposed chilled approximation. 
A related question is to understand how lowering the temperature leads to an acceleration of the convergence speed of the MCMC simulation.
Since the relative time step in our MCMC simulations appears to be roughly independent of temperature (for a given acceptance rate), a speculative answer is that the acceleration is exclusively associated with the slowest direction of the posterior distribution.\vspace{-0.2cm}\\

\noindent
\textbf{Acknowledgements.}\,\,\,
The authors wish to thank R\'egis Borde and Olivier Hautecoeur of the European operational satellite agency for monitoring of weather, climate and the environment from space (EUMETSAT) for providing the meteorological data and for fruitful discussions on operational assimilation issues in weather forecasting. \vspace{-0.cm}\\

\appendix 

\section{Simplified Atmospheric Dynamics}\label{app:modelMeteo}
By neglecting vertical winds and diabatic heating in the first law of thermodynamics expressed  in the isobaric system, we obtain the passive transport of temperature and specific humidity variables \cite{Holton92}. Moreover,   in atmosphere chemistry studies it is common to   treat ozone as a passive tracer below the stratosphere \cite{lahoz2010data}.   In consequence, using the isobaric system, \ie,  we can model the dynamic of any of these atmospheric variables denoted by $\tvt(q,p,t)\in  \mathcal{C}^1(\Omega \times \mathbb{R} \times \mathbb{R})$ taken on the points of  $\Omega$ and on specific pressure levels $p\in\mathbb R$  by the  simple transport \vspace{-0.2cm}
\begin{align}\label{eq:thermo}
 \frac{\partial \tvt}{\partial t}(q,p,t)  + \wwr(q,p,t) \cdot\nabla  \tvt(q,p,t) =0,
\end{align}
 parametrized by the horizontal wind field $\wwr(q,p,t)$ that verifies, at any time $t\in\mathbb R$ and pressure level $p\in\mathbb R$, $\wwr(.,p,t)\in L^2(\Omega)$.

 We are  interested in integrating \eqref{eq:thermo} vertically to relate the evolution of  pressure-averaged temperature, humidity or ozone fields to  pressure-averaged horizontal winds. We will use in the following a discretization of the continuous vertical pressure coordinate  into a finite set of decreasing pressure levels $\{p^k\}_{k=0}^K$  with $p^{k}>p^{k+1}$ for  $0\le k\le K-1$.
The  pressure-averaged fields are defined as:
$
\tvt^k(q,t)=\frac{1}{p^{k}-p^{k+1}}\int_{p^{k+1}}^{p^{k}} \tvt(q,p,t) dp.
$

In this perspective, we assume that horizontal winds  in the pressure  interval $[p^{k},p^{k+1}]$ are uniform and equal to pressure-average horizontal winds. Using  the Fubini theorem to invert derivatives and integrals,  vertical integration of ~\eqref{eq:thermo} yields a bi-dimensional transport equation of the form of~\eqref{eq:OFC}.

\section{Proof of lemma~\ref{lemma:approxsol}}\label{app:proofLemma1}\label{app:proofLemma}
In order to prove this lemma, we begin by characterizing the optimal solutions of the optimization problems \eqref{eq:minProb}.  \\

\begin{lemma}\label{lemma:sol}
The unique solution of problem \eqref{eq:minProb} with the  constraints \eqref{eq:const0} is
\begin{equation}\label{eq:weightL2}
{w}_{\mu,\boldsymbol{\hat \theta}}^{1}(\psi)= {  {  {{\mathcal{E}}_{\mu,\boldsymbol{\hat \theta}}(\psi)}^{-1}} }{\prod_{{\psi'}\in \mathcal{P}}{  {{\mathcal{E}}_{\mu,\boldsymbol{\hat \theta}}(\psi)}^{1/{\sharp\mathcal{P}(\Omega)}}}},\vspace{-0.2cm}
\end{equation}
while for the  constraints \eqref{eq:const1} \vspace{-0.2cm}
\begin{equation}\label{eq:weightL1}
{w}_{\mu,\boldsymbol{\hat \theta}}^{2}(\psi)=\left({{\sharp\mathcal{P}(\Omega)}} \frac{  {  {{\mathcal{E}}_{\mu,\boldsymbol{\hat \theta}}(\psi)}^{-1}} }{\sum_{{\psi'}\in \mathcal{P}}{  {{\mathcal{E}}_{\mu,\boldsymbol{\hat \theta}}(\psi)}^{-1}}}\right)^2,
\end{equation}
and finally the  constraints \eqref{eq:const2} yield 
 \begin{align}\label{eq:weightL0}
{w}_{\mu,\boldsymbol{\hat \theta}}^{0}(\psi)=
 \left\{\begin{aligned}
&{{\sharp\mathcal{P}(\Omega)}/\tau} \quad \textrm{if} \quad {  {  {{\mathcal{E}}_{\mu,\boldsymbol{\hat \theta}}(\psi)}} } \le  \textrm{c}\\
&0\quad \textrm{else} 
\end{aligned}\right.,
\end{align}
where the constant $ \textrm{c}\in [0,{\sum_{{\psi'}\in \mathcal{P}}{  {{\mathcal{E}}_{\mu,\boldsymbol{\hat \theta}}(\psi)}}}]$ is a function of $\tau$. \\
\end{lemma}

\proof{
The  solution \eqref{eq:weightL0} is trivial  noticing that the objective is a weighted sum of variance, with weights  either equal to the  positive constant ${{\sharp\mathcal{P}(\Omega)}/\tau}$ or to zero.  Indeed, considering $\tau$ non-zero elements,  the objective is obviously minimized by setting  non-zero weights to the $\tau$ elements $\psi \in  \mathcal{P}(\Omega)$ associated with the lowest expected error. 

Now, considering the change of variable  ${\tilde w}=\log{w}$ or  ${\tilde w}=\sqrt{{w}}$, the optimization problems \eqref{eq:minProb} subject to  \eqref{eq:const0} and \eqref{eq:const1}  are recasted in the minimization of a linear objective function subject to convex constraints.  According to~\cite[Proposition 5.2.1]{Bertsekas99} the solution of such constraint optimization problems can be characterized using a dual method, since there is no duality gap and there exists geometric multipliers. 

We  treat in parallel  the two optimization problems defined  either with the constraints \eqref{eq:const0} or \eqref{eq:const1}. Let us ignore in a first step the positive constraints related to  ${\tilde w}(\psi)\in \mathcal{W}$ and only consider the equality constraint. Denoting by $\lambda\in \Rr$  a geometric multiplier, the Lagrangian related to the problem  is 
\begin{align*}
\mathcal{L}({\tilde w}, \lambda)= \left\{\begin{aligned}
&\sum_{{\psi} \in \mathcal{P}(\Omega)} \exp^{{\tilde w}(\psi)}{{\mathcal{E}}_{\mu,\boldsymbol{\hat \theta}}(\psi)} - \lambda \sum_{{\psi} \in \mathcal{P}(\Omega)} {\tilde w}(\psi) \quad \textrm{for} \quad \eqref{eq:const0}\\
&\sum_{{\psi} \in \mathcal{P}(\Omega)} {\tilde w}^2(\psi){{\mathcal{E}}_{\mu,\boldsymbol{\hat \theta}}(\psi)} + \lambda( { {\sharp\mathcal{P}(\Omega)}}-\sum_{{\psi} \in \mathcal{P}(\Omega)} {\tilde w}(\psi) ) \quad  \textrm{for}  \quad \eqref{eq:const1}
\end{aligned}\right..
\end{align*}
The optimal solution of the problem of interest is the couple $({\tilde w}^\star(\psi), \lambda^\star)$ solution of \vspace{-0.2cm}
\begin{align*}
\max_{\lambda \in \Rr} \min_{{\tilde w} : \mathcal{P} \to \Rr} \mathcal{L}({\tilde w}, \lambda).
\end{align*}
Since the lagrangian is convex and concave respectively to  its first and second argument and the domain $\Rr$ is convex, the solution satisfies the first order optimal condition. We first cancel the derivatives with respect to ${\tilde w}$, which yields,  
\begin{align*}
\argmin_{{\tilde w} : \mathcal{P} \to \Rr} \mathcal{L}({\tilde w}, \lambda)
= \left\{\begin{aligned}
&\log(\lambda/{{\mathcal{E}}_{\mu,\boldsymbol{\hat \theta}}(\psi)}) \quad \textrm{for} \quad \eqref{eq:const0}\\
&-\frac{\lambda}{ 2{{\mathcal{E}}_{\mu,\boldsymbol{\hat \theta}}(\psi)}    } \quad  \textrm{for}  \quad \eqref{eq:const1}
\end{aligned}\right.,
\end{align*}
and canceling the derivative  with respect  to $\lambda$  of $ \min_{{\tilde w} : \mathcal{P} \to \Rr} \mathcal{L}({\tilde w}, \lambda)$ yields
\begin{align*}
{ \lambda^\star}= 
\left\{\begin{aligned}
{\prod_{\psi \in \mathcal{P}(\Omega)}{  {{\mathcal{E}}_{\mu,\boldsymbol{\hat \theta}}(\psi)}^{1/{\sharp\mathcal{P}(\Omega)}}}}\quad \textrm{for} \quad \eqref{eq:const0}\\
\frac{-2{\sharp\mathcal{P}(\Omega)}}{\sum_{\psi \in \mathcal{P}(\Omega)}{  {{\mathcal{E}}_{\mu,\boldsymbol{\hat \theta}}(\psi)}^{-1}}}\quad \textrm{for} \quad \eqref{eq:const1}\\
\end{aligned}\right.,
\end{align*}
and we finally obtain \vspace{-0.2cm}
\begin{align*}
{\tilde w}^\star(\psi)=
\left\{\begin{aligned}
{   {  {{\mathcal{E}}_{\mu,\boldsymbol{\hat \theta}}(\psi)}^{-1}}}{\prod_{\psi \in \mathcal{P}(\Omega)}{  {{\mathcal{E}}_{\mu,\boldsymbol{\hat \theta}}(\psi)}^{1/{\sharp\mathcal{P}(\Omega)}}}} \quad \textrm{for} \quad \eqref{eq:const0}\\
\frac{ {\sharp\mathcal{P}(\Omega)}  {  {{\mathcal{E}}_{\mu,\boldsymbol{\hat \theta}}(\psi)}^{-1}}}{\sum_{{\psi'}\in \mathcal{P}} {  {{\mathcal{E}}_{\mu,\boldsymbol{\hat \theta}}(\psi)}^{-1}}} \quad \textrm{for} \quad \eqref{eq:const1}\\
\end{aligned}\right..
\end{align*}
We then observe that ${\tilde w}^\star(\psi)>0$ for any $\psi\in \mathcal{P}$, \ie, already satisfies the positive constraints related to  ${\tilde w}(\psi)\in \mathcal{W}$. The sough solution follows by the reverse change of variable. $\square$\\
}

Now coming back to the proof of Lemma~\ref{lemma:approxsol}, it follows from the decomposition\vspace{-0.2cm}
\begin{align*}
\int \| \psi(\boldsymbol{\theta}^\star) - \psi({\boldsymbol{\hat \theta}})\|^2_2  \mu(d\boldsymbol{\theta}^\star)&=\int \| \psi(\boldsymbol{\theta}^\star) \|^2_2-2\psi(\boldsymbol{\theta}^\star)^\intercal\psi({\boldsymbol{\hat \theta}}) +\|\psi({\boldsymbol{\hat \theta}})\|_2^2+\|\psi(\boldsymbol{\theta}_{PM})\|_2^2-\|\psi(\boldsymbol{\theta}_{PM})\|_2^2  \mu(d\boldsymbol{\theta}^\star),\\
&= \int \|  \psi(\boldsymbol{\theta}^\star)-\psi(\boldsymbol{\theta}_{PM}) \|^2_2+ \|  \psi({\boldsymbol{\hat \theta}})-\psi(\boldsymbol{\theta}_{PM}) \|^2_2  \mu(d\boldsymbol{\theta}^\star),
\end{align*}
and from the definitions of the ${\textrm{EPE}}$ that $\int \textrm{EPE}(\Omega,{w}_{\hat \mu,{\boldsymbol{\hat \theta}}}^{p},\boldsymbol{\theta^\star}-  {\boldsymbol{\hat \theta}})  \mu(d\boldsymbol{\theta}^\star)=$\vspace{-0.2cm}
\begin{align*}
& \frac{1}{\sharp\mathcal{P}(\Omega)}\sum_{{\psi} \in \mathcal{P}(\Omega)} {w}_{\hat \mu,{\boldsymbol{\hat \theta}}}^{p}(\psi) \int \left( \|  \psi(\boldsymbol{\theta}^\star)-\psi(\boldsymbol{\theta}_{PM}) \|^2_2+ \|  \psi({\boldsymbol{\hat \theta}})-\psi(\boldsymbol{\theta}_{PM}) \|^2_2\right)^{1/2} \mu(d\boldsymbol{\theta}^\star),\\
&\ge\int  \textrm{EPE}(\Omega,{w}_{\hat \mu,{\boldsymbol{\hat \theta}}}^{p},\boldsymbol{\theta^\star}-  \boldsymbol{\theta}_{PM}) \mu(d\boldsymbol{\theta}^\star)\ge \int \textrm{EPE}(\Omega,{w}_{\mu,{\boldsymbol{\theta}}_{PM}}^{p},\boldsymbol{\theta^\star}-  \boldsymbol{\theta}_{PM}) \mu(d\boldsymbol{\theta}^\star),
\end{align*}
where the last inequality follows from Lemma~\ref{lemma:sol}. $\quad\quad\quad\quad\quad\quad\quad\quad\quad\quad\quad\quad\quad\quad \square$

\section{Variable Decomposition, Gradient and Hessian}\label{app:fastGrad}

\subsection{Fourier-Wavelet  fBm Representation}\label{app:fBm} 
  It can be shown following the lines of proof of  \cite[Proposition 3.1]{Heas14} that under mild conditions   an isotropic fBm  admits the wavelet representation\vspace{-0.2cm}
\begin{align}\label{eq:fBmRepWav}
\sum_{j \in \Omega_j} {a}_j {\psi}^{(-H-1)}_{{ j}}(\mathbf{s}), \quad \forall \mathbf{s}\in \Omega,
\end{align} 
where $\Omega_j$ is the (infinite) set of indices so that $ \{{\psi}_{{ j}}\}_j$ forms an orthonormal basis of $L^2(\Omega)$,  where  elements in the set $\{a_j\}_j$ are  i.i.d.  random variable distributed according to the   standard normal law, and  where the fractional Laplacian of a $\lceil \xi \rceil $ times differentiable wavelet denoted formally by ${\psi}^{\xi}_{i}(\mathbf{s}) \triangleq (-\Delta)^{\frac{\xi}{2}}{\psi}_i$ is  defined in the Fourier domain. 

Now, let the orthonormal subset  $\{\psi_i\}_{i=1}^m$  be chosen so that  wavelet coefficients revealing scales smaller than the pixel size will
be neglected.   Moreover, assume an interpolant and anisotropic  wavelet basis, such as for instance  Coiflets.
 Truncating the  wavelet series~\eqref{eq:fBmRepWav},  we define independent  random fields  $\boldsymbol{ \theta}^\ell\in \Rr^m$ on the pixel grid $\Omega_m$ with $l=1,\ldots,k+2$. These random fields, approximating isotropic fBms, may be noted in the  form of the matrix-vector products  $\boldsymbol{ \theta}^\ell=\boldsymbol{\Psi}_m^{-H-1}\mathbf{a}^{\ell}$, with matrix $\boldsymbol{\Psi}_m^{-H-1} \in \Rr^{m \times m}  $ and the coefficient vector $\mathbf{a}^{\ell} \in \Rr^m$.  
 
We show hereafter that for sufficiently large $m$, the covariance of the finite random-field $\boldsymbol{ \theta}^\ell$ approximates accurately the covariance function of an isotropic fBm.  Moreover, there exists a tractable approximation of its inverse. Indeed, let components of $\boldsymbol{ \theta}^\ell \in \Rr^{m}$ be zero-mean correlated   Gaussian random variables of covariance matrix  $\boldsymbol{\Psi}_m  {A}_H  \boldsymbol{\Psi}_m^\intercal$, where the entries of $A_H\in \Rr^{m \times m}$ are given  for any $i, i' =1,\ldots,m$   by\vspace{-0.2cm}
\begin{equation*}
 A_H(i,i') =  
 \langle  {\psi}^{(-H-1)}_{{ i}} , {\psi}^{(-H-1)}_{{i'}} \rangle_{L^2(\Omega)},
 \end{equation*}
 and where we denote the inverse wavelet transform of a two-dimensional random field in $\Rr^m$ using  the orthonormal basis $\{\psi_i\}_{i=1}^m$  by  $\boldsymbol{\Psi}_m$. \\
 
\begin{proposition} \label{prop:covDls}
 Assume that $\sum_{j=1}^{m}\boldsymbol{ \theta}^\ell (j)=0$.  Then,
 in the limit of $m\to \infty$, the entries of  matrix $\boldsymbol{\Psi}_m  {A}_H  \boldsymbol{\Psi}_m^\intercal$ is the covariance function of a zero-mean  isotropic fBm  of parameter $H$,
and moreover, 
 the components  of the inverse of matrix  $A_H$  are
$$
 A_H^{-1}(i,i') =  
 \langle  {\psi}^{(H+1)}_{{ i}} , {\psi}^{(H+1)}_{{i'}} \rangle_{L^2(\Omega)}.
 $$\\
 \end{proposition}
This result  is shown by  following the lines of the proof of \cite[Proposition 3.2]{Heas14} and \cite[Lemma 4.1]{Heas14}.

We detail hereafter how to compute efficiently the gradient and the Hessian of the log posterior. 
\subsection{Fast gradient computation}\label{sec:gradient}
 Some  analytical calculations leads to the gradients  at $\boldsymbol{\theta}=(\mathbf{d}^\intercal,{\xx}_{t_1}^\intercal)^\intercal$ given by
 \begin{align*}
-\zeta' \nabla_{{\mathbf{d}}} \log \mu(\boldsymbol{\theta})&=\alpha' \left( \begin{pmatrix}\boldsymbol{ \Psi}_m \\ 0 \end{pmatrix} {A}_H^{-1} \begin{pmatrix}\boldsymbol{ \Psi^\intercal}_m & 0 \end{pmatrix} + \begin{pmatrix} 0 \\\boldsymbol{ \Psi}_m \end{pmatrix} {A}_H^{-1} \begin{pmatrix}0 &\boldsymbol{ \Psi^\intercal}_m \end{pmatrix} \right) \mathbf{d}  \\
&+ \nabla_{\mathbf{d}} \mathcal{W}^\intercal(\xx_{t_1},\mathbf{d})\, {\boldsymbol{\delta}_{t_0}(\boldsymbol{\theta})},\\
-\zeta'  \nabla_{\xx_{t_1}}  \log \mu(\boldsymbol{\theta})&=  {\gamma}' \xx_{t_1}  +
 \nabla_{{\xx}} \mathcal{W}^\intercal(\xx_{t_1},\mathbf{d}){\boldsymbol{\delta}_{t_0}}(\boldsymbol{\theta})+{\boldsymbol{\delta}_{t_1}(\boldsymbol{\theta})}.
 \end{align*} 

The methodology for the gradient evaluation is analogous to the one proposed in~\cite{heas2016efficient}.
First,  we begin by making some comments  on the evaluation of the warping   defined in~\eqref{eq:splineRepr}.
 We propose to use   the family of bi-dimensional cubic cardinal   splines  $\{\varphi_i\}_{i=1}^{km}$ for their representation.
In practice, we compute an equivalent representation  based on   the family of bi-dimensional  cubic  B-splines functions $\{\phi_i\}_{i=1}^{km}$. Indeed,  this  representation presents some computational advantages because of the existence of fast B-splines transforms. The relation between cardinal  cubic splines and cubic  B-splines functions is given in \cite{Unser91}. This reference also provides details on the fast cubic  B-splines transform by recursive filtering.  Let matrix $\mathbf{C}^\intercal=[\mathbf{c}_p,...,\mathbf{c}_n]^\intercal \in\Rr^{km \times km}$ be the direct B-spline transform  of $k$ discrete bi-dimensional signal, \ie, the transform computing from the vector $\xx_{t_1}$ its representation with spline coefficients $ \mathbf{C}^\intercal\xx_{t_1}$.
To simplify notations, we denote by $\mathcal{I}:\Rr^{km} \times \Rr^{2m} \to \Rr^{km}$ the function taking as a first argument spline coefficients   $\mathbf{C}^\intercal\xx_{t_1}$ and as a second argument a motion field $\mathbf{d}$, and whose components are  given by  $\mathcal{W}_s(\xx_{t_1},\mathbf{d})$  defined in~\eqref{eq:splineRepr}. 
   Using this  notation,  the vector 
$
   \mathcal{I}(\mathbf{C}^\intercal\xx_{t_1},\mathbf{d}),
$
 has its $s$-th component  given by 
$
 \sum_{i \in \vartheta(\varkappa({s})+ \mathbf{d}({s}))} \mathbf{c}_i^\intercal\xx_{t_1} \, \phi_i(\varkappa({s})+ \mathbf{d}({s})),  
$
   with   $\vartheta(\varkappa(s)+\mathbf{d}({s}))$ denoting a subset of function index  in the neighborhood of the spatial position $\varkappa({s})$. 
 Now, we denote by   $\nabla \mathcal{I}( \mathbf{C}^\intercal\xx_{t_1} ,\mathbf{d})$ the Jacobian of function $\mathcal{I}$ at point $(\mathbf{C}^\intercal\xx_{t_1} ,\mathbf{d})$ with respect to its first argument, \ie, spline coefficients.  Since  function $\mathcal{I}$ is linear with respect to  spline coefficients, the Jacobian is only dependent on the value of its second argument, \ie, $\mathbf{d}$. Therefore, we will adopt  the notation $\nabla \mathcal{I}(\mathbf{d})$ in the sequel\footnote{Note that operator $\nabla_{{\xx}}\mathcal{W}^\intercal({\xx},\mathbf{d})$ may be written $\nabla_{{\xx}}\mathcal{W}^\intercal(\mathbf{d})$ as well,  as it is only dependent on the value of its second argument, \ie, $\mathbf{d}$}.  
The complexity of evaluating both spline coefficients $\mathbf{C}^\intercal\xx_{t_1} $ and  the interpolated function $\mathcal{I}$, scales linearly with the image dimension, \ie, $\mathcal{O}(n)$, thanks to the representation separability and to recursive linear filtering \cite{Unser91}. Multiplication of $\mathbf{C}$  with the Jacobian   transpose, we get the Jacobian transpose of the warping function 
$
 \nabla_{{\xx}}\mathcal{W}^\intercal({\xx},\mathbf{d})=\mathbf{C} \nabla \mathcal{I}^\intercal(\mathbf{d}),
 $
 which also implies a linear complexity: first,  matrix $\mathbf{C}$ is symmetric\footnote{Matrix $\mathbf{C}$ is symmetric in the case of periodic boundary conditions  \cite{Unser91}.} 
  so that it is identical to  the direct B-spline transformation  $\mathbf{C}^\intercal $, computed by recursive linear filtering; second, the  multiplication of the Jacobian transpose  of function $\mathcal{I}$  with a vector $\mathbf{z}\in \Rr^{km}$ has its $s$-th component equal to
  $\nabla \mathcal{I}^\intercal(\mathbf{d}) \mathbf{z}({s})=\sum_{\mathbf{i} | {s} \in \vartheta(\varkappa({i})+\mathbf{d}({i}))}  \mathbf{z} ({i}) \phi_s(\varkappa({i})+\mathbf{d}({i})).$  
%
%
%
%
%

Second, let  $\mathbf{d}_i$ with $i=1,2$ be the two components of the  bivariate displacement field, and $\boldsymbol{ \Psi^\intercal}_m  \mathbf{d}_i=\mathbf{e}_{\mathbf{d}_i}$ be the vector  of the related wavelet coefficients.  We remark that 
the components of  $\boldsymbol{ \Psi}{A}_H^{-1}\boldsymbol{ \Psi^\intercal}_m  \mathbf{d}_i$ can be evaluated for $i=1,2$  at once by  computing the fast Fourier transform (FFT) of  $\mathbf{d}_i$, then applying the fractional differentiation operator in the Fourier domain, and finally performing an inverse FFT.
  
 Finally, the   matrix $\nabla_{\mathbf{d}} \mathcal{W}^\intercal(\xx_{t_1},\mathbf{d})$ is composed of two-by-$k$  blocks of size $m$ times $m$, which are all diagonal. The  components on the diagonal of the $k$ upper blocks (resp. the $k$ lower blocks) are the components of the $n$-dimensional vector   $\partial_{s_j} \left(\mathcal{W}^\intercal(\xx_{t_1},\mathbf{d})\right) $ for $j=1$ (resp. for $j=2$),   where   $s
 _j$ denotes the $j$-th spatial coordinate.  
We approach  these partial derivatives by second-order centered finite differences.  

  The overall complexity for evaluating the gradients  is thus  bounded by the FFT computation in $\mathcal{O}(m\log m)$.

\subsection{Laplace approximation in practice}\label{rem:CondIndLaplace}
In the perspective of computing a Laplace approximation, we need the components of the Hessian matrix, which are given for $ s,s'\in \{1,\ldots,m\}$,  for $ r,r'\in \{1,\ldots,n\}$ and $i,i'\in\{1,2\}$ as:
\begin{align*}
-\zeta' \frac{\partial^2 \log \mu(\boldsymbol{\theta})}{\partial \mathbf{d}_i(s)\partial\mathbf{d}_{i'}(s')}&=
\delta_{i=i'}  \alpha' \left( \boldsymbol{ \Psi}_m {A}_H^{-1}\boldsymbol{ \Psi^\intercal}_m \right)_{(s,s')} \\
 +\delta_{s=s'} &
 \begin{pmatrix}
 \partial^2_{s_i,s_{i'}}   \mathcal{W}_{s}( \xx_{t_1},\mathbf{ d}) {\boldsymbol{\delta}_{t,s}}(\boldsymbol{\theta})+ \partial_{s_i}   \mathcal{W}_{s}( \xx_{t_1},\mathbf{ d}) \partial_{s_{i'}}   \mathcal{W}_{s}( \xx_{t_1},\mathbf{ d})\unit_{s\in \Omega_{obs}^{t}}
 \end{pmatrix},\\
-\zeta' \frac{\partial^2 \log \mu(\boldsymbol{\theta})}{\partial \mathbf{d}_i(s) \partial\xx(r')}&=
  \partial_{\mathbf{d}_i(s)} \{ \begin{pmatrix}
 \nabla_{\xx} \mathcal{W}^\intercal(\mathbf{ d})  {\boldsymbol{\delta}_{t,s}}(\boldsymbol{\theta})
 \end{pmatrix}(r')\},\\
-\zeta' \frac{\partial^2 \log \mu(\boldsymbol{\theta})}{\partial \xx(r) \partial\xx(r')}&=
  \begin{pmatrix}
 \nabla_{\xx} \mathcal{W}^\intercal(\mathbf{ d}) \diag(\unit_{ \Omega_{obs}^{t} })\nabla_{\xx} \mathcal{W}(\mathbf{ d}) + \diag(\unit_{ \Omega_{obs}^{t_1}}) 
 \end{pmatrix}_{(r,r')}+ \gamma' \delta_{r=r'} ,
\end{align*}

The elements  $ \{\left( \boldsymbol{ \Psi}_m {A}_H^{-1}\boldsymbol{ \Psi^\intercal}_m \right)_{(s,s')}\}_{s=1}^m$ for an arbitrary $s'$ are evaluated by computing the product of the matrix  $\boldsymbol{ \Psi}_m {A}_H^{-1}\boldsymbol{ \Psi^\intercal}$  with the vector with a non-zero unit entry at $s'=s$, thanks to  the use of the direct and inverse FFT. Due to translation invariance on the two-dimensional pixel grid, the $\boldsymbol{ \Psi}_m {A}_H^{-1}\boldsymbol{ \Psi^\intercal}$ matrix is  diagonal band, with the absolute value of the matrix components becoming negligible far from the diagonal. In practice we use an approximation of this diagonal band matrix: we remove negligible dependences in order to restrict the dimension of the neighborhood and accelerate computation of the Laplace approximation, as we will discuss in the next section. The other terms in the second order derivative with respect to displacement are approximated by finite differences. 
The term $ \partial_{\mathbf{d}_i(s)} \{ \begin{pmatrix}
 \nabla_{\xx} \mathcal{W}^\intercal(\mathbf{ d})  {\boldsymbol{\delta}_{t,s}}(\boldsymbol{\theta})
 \end{pmatrix}(r')\}$ is approximated from 
  $ \nabla_{\xx} \mathcal{W}^\intercal(\mathbf{ d})  {\boldsymbol{\delta}_{t_0}}(\boldsymbol{\theta})$ by finite differences and assuming a locally constant displacement.  Finally, the non-zero elements $\begin{pmatrix}
 \nabla_{\xx} \mathcal{W}^\intercal(\mathbf{ d}) \diag(\unit_{ \Omega_{obs}^{t} })\nabla_{\xx} \mathcal{W}(\mathbf{ d}) 
 \end{pmatrix}_{(r,r')}$ are explicitly computed.

The complexity to compute the Hessian and compute its EVD  scales as $\mathcal{O}(n^3)$. It can be significantly reduced exploiting conditional independence
 between subsets of components  of  vector $\boldsymbol{\theta}$. More precisely, consider the  decomposition $\boldsymbol{\theta}=(\boldsymbol{\theta}_1^\intercal,\boldsymbol{\theta}_2^\intercal,\boldsymbol{\theta}_3^\intercal)^\intercal$ with $\boldsymbol{\theta}_1\in\mathbf{\Theta_1}$, $\boldsymbol{\theta}_2\in\mathbf{\Theta_2}$ and $\boldsymbol{\theta}_3\in\mathbf{\Theta_3}$, and
where the $j$-th component of variable $\boldsymbol{\theta}_1$ is $\mathbf{\psi}_j^\intercal\boldsymbol{ \theta}$. It is straightforward to see that if   $\boldsymbol{\theta}_1$  is   independent from  $\boldsymbol{\theta}_3$ conditionally to neighborhood variables $\boldsymbol{\theta}_2$,
 then
 it follows that the bound in ~\eqref{eq:boundEllipsRestrict} can be rewritten as 
 $\mathcal{F}=\sqrt{\frac{2}{\pi}}   \sum_{j=1}^\ell \| \Lambda_{{\mathbf{H}'_{U}}}^{-1/2} \mathbf{V}_{{\mathbf{H}'_{U}}}^\intercal   {\mathbf{\psi}'_j}\|_2,$ 
where  $ \mathbf{\psi}'_j$ is the restriction of $\mathbf{\psi}_j$ on $\mathbf{\Theta_1}\cup\mathbf{\Theta_2}$ and where  $\Lambda_{\mathbf{H}'_{U}}$'s and $\mathbf{V}_{\mathbf{H}'_{U}}$'s are the diagonal eigenvalue matrix and the matrix whose columns are the eigenvectors of the covariance matrix related to the marginal
$\int_{\mathbf{\Theta_3}} \mu(d{\boldsymbol{\theta}})$.  In the case where $n'=\dim(\mathbf{\Theta_1}\cup\mathbf{\Theta_2}) \ll n$,  the complexity to compute the error bound estimate $\mathcal{F}$ is significantly  reduced to  $\mathcal{O}(n'^3)$.
The dimension of the neighborhood\footnote{
The neighborhood  $ \partial_{\boldsymbol{\theta}(i)}$  of the $i$-th component $  \boldsymbol{\theta}(i)$ is  defined as the set  of $ \boldsymbol{\theta}(j)$ with $j\neq i$ such that for any with $k\neq i$,  components $ \boldsymbol{\theta}(i)$ and  $\boldsymbol{\theta}(k) $  are conditionally independent with respect to  $\partial_{\boldsymbol{\theta}(i)}$.
 Under our Gaussian approximation,
  the neighborhood  $ \partial_{\boldsymbol{\theta}(i)}$  is the set  of $ \boldsymbol{\theta}(j)$ with $j\neq i$ such that   the $(i,j)$-th entry of matrix ${\mathbf{H}_{U}}$  is not equal to zero.
%
 }
 is in general much smaller than $n$. Indeed, many  entries of  ${\mathbf{H}_{U}}$, related to variables located sufficiently far away, are equal to zero or can be neglected. This  property often characterizes  random fields, and has been extensively exploited in  image analysis~\cite{wang2013markov}. 

\bibliographystyle{spmpsci}
{
\bibliography{ref,group-15302,cherzet,old}

\begin{thebibliography}{10}
\providecommand{\url}[1]{{#1}}
\providecommand{\urlprefix}{URL }
\expandafter\ifx\csname urlstyle\endcsname\relax
  \providecommand{\doi}[1]{DOI~\discretionary{}{}{}#1}\else
  \providecommand{\doi}{DOI~\discretionary{}{}{}\begingroup
  \urlstyle{rm}\Url}\fi

\bibitem{berger2013statistical}
Berger, J.O.: Statistical decision theory and Bayesian analysis.
\newblock Springer Science \& Business Media (2013)

\bibitem{Bertsekas99}
Bertsekas, D.P.: Nonlinear Programming, 2nd edn.
\newblock Athena Scientific (1999)

\bibitem{beskos2017geometric}
Beskos, A., Girolami, M., Lan, S., Farrell, P.E., Stuart, A.M.: Geometric mcmc
  for infinite-dimensional inverse problems.
\newblock Journal of Computational Physics \textbf{335}, 327--351 (2017)

\bibitem{beskos2009mcmc}
Beskos, A., Stuart, A.: Mcmc methods for sampling function space.
\newblock In: 6th International Congress on Industrial and Applied Mathematics,
  pp. 337--364. Citeseer (2009)

\bibitem{borde2019winds}
Borde, R., Carranza, M., Hautecoeur, O., Barbieux, K.: Winds of change for
  future operational amv at eumetsat.
\newblock Remote Sensing \textbf{11}(18), 2111 (2019)

\bibitem{Butler:ECCV:2012}
Butler, D.J., Wulff, J., Stanley, G.B., Black, M.J.: A naturalistic open source
  movie for optical flow evaluation.
\newblock In: {A. Fitzgibbon et al. (Eds.)} (ed.) European Conf. on Computer
  Vision (ECCV), Part IV, LNCS 7577, pp. 611--625. Springer-Verlag (2012)

\bibitem{cotter2013mcmc}
Cotter, S.L., Roberts, G.O., Stuart, A.M., White, D.: Mcmc methods for
  functions: modifying old algorithms to make them faster.
\newblock Statistical Science \textbf{28}(3), 424--446 (2013)

\bibitem{derian2013wavelets}
D{\'e}rian, P., H{\'e}as, P., Herzet, C., M{\'e}min, E.: Wavelets and optical
  flow motion estimation.
\newblock Numerical Mathematics: Theory, Methods and Applications
  \textbf{6}(1), 116--137 (2013)

\bibitem{girolami2011riemann}
Girolami, M., Calderhead, B.: Riemann manifold langevin and hamiltonian monte
  carlo methods.
\newblock Journal of the Royal Statistical Society: Series B (Statistical
  Methodology) \textbf{73}(2), 123--214 (2011)

\bibitem{hastings1970monte}
Hastings, W.K.: Monte carlo sampling methods using markov chains and their
  applications  (1970)

\bibitem{heas2016efficient}
H{\'e}as, P., Dr{\'e}meau, A., Herzet, C.: An efficient algorithm for video
  superresolution based on a sequential model.
\newblock SIAM Journal on Imaging Sciences \textbf{9}(2), 537--572 (2016)

\bibitem{heas2011bayesian}
H{\'e}as, P., Herzet, C., M{\'e}min, E.: Bayesian inference of models and
  hyperparameters for robust optical-flow estimation.
\newblock IEEE Transactions on Image Processing \textbf{21}(4), 1437--1451
  (2011)

\bibitem{heas2012bayesian}
H{\'e}as, P., Herzet, C., M{\'e}min, E., Heitz, D., Mininni, P.D.: Bayesian
  estimation of turbulent motion.
\newblock IEEE transactions on pattern analysis and machine intelligence
  \textbf{35}(6), 1343--1356 (2012)

\bibitem{Heas14}
H{\'e}as, P., Lavancier, F., Harouna, S.K.: {Self-similar prior and wavelet
  bases for hidden incompressible turbulent motion}.
\newblock {SIAM Journal on Imaging Sciences} \textbf{{7}}({2}), {1171--1209,}
  ({2014})

\bibitem{Holton92}
Holton, J.: An introduction to dynamic meteorology.
\newblock Academic press (1992)

\bibitem{Heas_2023}
Héas, P., Hautecoeur, O., Borde, R.: 3d wind field profiles from hyperspectral
  sounders: revisiting optic-flow from a meteorological perspective.
\newblock Physica Scripta \textbf{98}(11), 115,208 (2023)

\bibitem{ilg2018uncertainty}
Ilg, E., Cicek, O., Galesso, S., Klein, A., Makansi, O., Hutter, F., Brox, T.:
  Uncertainty estimates and multi-hypotheses networks for optical flow.
\newblock In: Proceedings of the European Conference on Computer Vision (ECCV),
  pp. 652--667 (2018)

\bibitem{Kadri13}
Kadri~Harouna, S., D{\'e}rian, P., H{\'e}as, P., M{\'e}min, E.:
  {Divergence-free Wavelets and High Order Regularization}.
\newblock International Journal of Computer Vision \textbf{103}(1), 80--99
  (2013)

\bibitem{kondermann2007adaptive}
Kondermann, C., Kondermann, D., J{\"a}hne, B., Garbe, C.: An adaptive
  confidence measure for optical flows based on linear subspace projections.
\newblock In: Joint Pattern Recognition Symposium, pp. 132--141. Springer
  (2007)

\bibitem{kondermann2008statistical}
Kondermann, C., Mester, R., Garbe, C.: A statistical confidence measure for
  optical flows.
\newblock In: European Conference on Computer Vision, pp. 290--301. Springer
  (2008)

\bibitem{krajsek2006maximum}
Krajsek, K., Mester, R.: A maximum likelihood estimator for choosing the
  regularization parameters in global optical flow methods.
\newblock In: 2006 International Conference on Image Processing, pp.
  1081--1084. IEEE (2006)

\bibitem{kybic2011bootstrap}
Kybic, J., Nieuwenhuis, C.: Bootstrap optical flow confidence and uncertainty
  measure.
\newblock Computer Vision and Image Understanding \textbf{115}(10), 1449--1462
  (2011)

\bibitem{lahoz2010data}
Lahoz, W., Khattatov, B., Menard, R.: Data Assimilation: Making Sense of
  Observations.
\newblock Springer Berlin Heidelberg (2010)

\bibitem{le2004error}
Le~Marshall, J., Rea, A., Leslie, L., Seecamp, R., Dunn, M.: Error
  characterisation of atmospheric motion vectors.
\newblock Australian Meteorological Magazine \textbf{53}(2) (2004)

\bibitem{Liu08}
{Liu}, T., {Shen}, L.: {Fluid flow and optical flow}.
\newblock Journal of Fluid Mechanics \textbf{614}, 253--291 (2008)

\bibitem{luan2020langevin}
Luan, F., Zhao, S., Bala, K., Gkioulekas, I.: Langevin monte carlo rendering
  with gradient-based adaptation.
\newblock ACM Trans. Graph. \textbf{39}(4), 140 (2020)

\bibitem{mac2012learning}
Mac~Aodha, O., Humayun, A., Pollefeys, M., Brostow, G.J.: Learning a confidence
  measure for optical flow.
\newblock IEEE transactions on pattern analysis and machine intelligence
  \textbf{35}(5), 1107--1120 (2012)

\bibitem{metropolis1953equation}
Metropolis, N., Rosenbluth, A.W., Rosenbluth, M.N., Teller, A.H., Teller, E.:
  Equation of state calculations by fast computing machines.
\newblock The journal of chemical physics \textbf{21}(6), 1087--1092 (1953)

\bibitem{neal2011mcmc}
Neal, R.M., et~al.: Mcmc using hamiltonian dynamics.
\newblock Handbook of markov chain monte carlo \textbf{2}(11), 2 (2011)

\bibitem{pereyra2015survey}
Pereyra, M., Schniter, P., Chouzenoux, E., Pesquet, J.C., Tourneret, J.Y.,
  Hero, A.O., McLaughlin, S.: A survey of stochastic simulation and
  optimization methods in signal processing.
\newblock IEEE Journal of Selected Topics in Signal Processing \textbf{10}(2),
  224--241 (2015)

\bibitem{phillpot1992temperature}
Phillpot, S., Rickman, J.: Temperature dependence of the thermodynamic
  properties of a liquid over a wide range of temperatures from simulations at
  a single temperature.
\newblock Molecular Physics \textbf{75}(1), 189--195 (1992)

\bibitem{raviart1983introduction}
Raviart, P., Thomas, J.: Introduction {\`a} l'analyse num{\'e}rique des
  {\'e}quations aux d{\'e}riv{\'e}es partielles.
\newblock Collection Math{\'e}matiques appliqu{\'e}es pour la ma{\^\i}trise.
  Masson (1983)

\bibitem{rickman1991temperature}
Rickman, J., Phillpot, S.: Temperature dependence of thermodynamic quantities
  from simulations at a single temperature.
\newblock Physical review letters \textbf{66}(3), 349 (1991)

\bibitem{roberts2004general}
Roberts, G.O., Rosenthal, J.S.: General state space markov chains and mcmc
  algorithms.
\newblock Probability surveys \textbf{1}, 20--71 (2004)

\bibitem{roberts1996exponential}
Roberts, G.O., Tweedie, R.L.: Exponential convergence of langevin distributions
  and their discrete approximations.
\newblock Bernoulli pp. 341--363 (1996)

\bibitem{santek20192018}
Santek, D., Dworak, R., Nebuda, S., Wanzong, S., Borde, R., Genkova, I.,
  Garc{\'\i}a-Pereda, J., Galante~Negri, R., Carranza, M., Nonaka, K., et~al.:
  2018 atmospheric motion vector (amv) intercomparison study.
\newblock Remote Sensing \textbf{11}(19), 2240 (2019)

\bibitem{santek2019demonstration}
Santek, D., Nebuda, S., Stettner, D.: Demonstration and evaluation of 3d winds
  generated by tracking features in moisture and ozone fields derived from airs
  sounding retrievals.
\newblock Remote Sensing \textbf{11}(22), 2597 (2019)

\bibitem{stoll2017time}
Stoll, M., Volz, S., Maurer, D., Bruhn, A.: A time-efficient optimisation
  framework for parameters of optical flow methods.
\newblock In: Scandinavian Conference on Image Analysis, pp. 41--53. Springer
  (2017)

\bibitem{sun2018bayesian}
Sun, J., Quevedo, F.J., Bollt, E.: Bayesian optical flow with uncertainty
  quantification.
\newblock Inverse Problems \textbf{34}(10), 105,008 (2018)

\bibitem{Suter94}
Suter, D.: Motion estimation and vector splines.
\newblock In: Proc. Conf. Comp. Vision Pattern Rec., pp. 939--942. Seattle, USA
  (1994)

\bibitem{Tafti11}
Tafti, P.D., Unser, M.: On regularized reconstruction of vector fields.
\newblock Image Processing, IEEE Trans. on \textbf{20}(11), 3163 --3178 (2011)

\bibitem{teixeira2021using}
Teixeira, J.V., Nguyen, H., Posselt, D.J., Su, H., Wu, L.: Using machine
  learning to model uncertainty for water vapor atmospheric motion vectors.
\newblock Atmospheric Measurement Techniques \textbf{14}(3), 1941--1957 (2021)

\bibitem{temperton2001two}
Temperton, C., Hortal, M., Simmons, A.: A two-time-level semi-lagrangian global
  spectral model.
\newblock Quarterly Journal of the Royal Meteorological Society
  \textbf{127}(571), 111--127 (2001)

\bibitem{ummenhofer2017demon}
Ummenhofer, B., Zhou, H., Uhrig, J., Mayer, N., Ilg, E., Dosovitskiy, A., Brox,
  T.: Demon: Depth and motion network for learning monocular stereo.
\newblock In: Proceedings of the IEEE conference on computer vision and pattern
  recognition, pp. 5038--5047 (2017)

\bibitem{Unser91}
Unser, M., Aldroubi, A., Eden, M.: Fast \mbox{{B}-Spline} transforms for
  continuous image representation and interpolation.
\newblock {IEEE} Transactions on Pattern Analysis and Machine Intelligence
  \textbf{13}(3), 277--285 (1991)

\bibitem{van1987simulated}
Van~Laarhoven, P.J., Aarts, E.H.: Simulated annealing.
\newblock In: Simulated annealing: Theory and applications, pp. 7--15. Springer
  (1987)

\bibitem{wang2013markov}
Wang, C., Komodakis, N., Paragios, N.: Markov random field modeling, inference
  \& learning in computer vision \& image understanding: A survey.
\newblock Computer Vision and Image Understanding \textbf{117}(11), 1610--1627
  (2013)

\bibitem{wannenwetsch2017probflow}
Wannenwetsch, A.S., Keuper, M., Roth, S.: Probflow: Joint optical flow and
  uncertainty estimation.
\newblock In: Proceedings of the IEEE international conference on computer
  vision, pp. 1173--1182 (2017)

\bibitem{wright1999numerical}
Wright, S., Nocedal, J., et~al.: Numerical optimization.
\newblock Springer Science \textbf{35}(67-68), 7 (1999)

\end{thebibliography}
}

\end{document}